\documentclass[letterpaper,12pt,peerreviewca,draftcls]{Resources/IEEEtran}
\usepackage{Resources/csm16}
\usepackage[margin=1in]{geometry}
\usepackage{amsmath} 

\usepackage{multirow}

\usepackage{mathrsfs}

\usepackage{url}
\usepackage{graphicx,xcolor}
\usepackage{verbatim}
\makeatletter
\newcommand{\verbatimfont}[1]{\def\verbatim@font{#1}}%
\makeatother
\verbatimfont{\ttfamily\small}

\newcommand{\bi}{\begin{itemize}}\newcommand{\ei}{\end{itemize}}
\newcommand{\be}{\begin{equation}}\newcommand{\ee}{\end{equation}}
\newcommand{\bee}{\begin{enumerate}}\newcommand{\eee}{\end{enumerate}}
\newcommand{\bea}{\begin{eqnarray}}\newcommand{\eea}{\end{eqnarray}}
\newcommand{\beas}{\begin{eqnarray*}}\newcommand{\eeas}{\end{eqnarray*}}
\newcommand{\bc}{\begin{center}}\newcommand{\ec}{\end{center}}
\usepackage[left,pagewise]{lineno} 
\usepackage[english]{babel} 
\usepackage{blindtext}

\usepackage{amsmath}
\usepackage{amssymb}
\usepackage{amsfonts}

\usepackage{amsthm}
\usepackage{comment}
\usepackage{cite}

\renewcommand\thefigure{\arabic{figure}}

\theoremstyle{plain}

\theoremstyle{plain}

\theoremstyle{definition}

\theoremstyle{definition}

\theoremstyle{definition}

\usepackage[framemethod=TikZ]{mdframed}
\usepackage{adjustbox}
\usepackage[most]{tcolorbox}
\tcbset{enhanced jigsaw,colback=blue!5!white,colframe=blue!55!green,
fonttitle=\bfseries}

\title{
Mechanism Design Theory in Control Engineering \\
\Large A Tutorial and Overview of Applications in Communication, Power Grid, Transportation, and Security Systems}
\author{Ioannis Vasileios Chremos, and Andreas A. Malikopoulos\\
	POC: A. A.\ Malikopoulos (andreas@udel.edu)\\ \today }

\newif\ifPDF \ifx\pdfoutput\undefined\PDFfalse \else\ifnum\pdfoutput > 0\PDFtrue \else\PDFfalse \fi \fi
\ifPDF 
\usepackage[pdftex, plainpages = false, colorlinks=true, linkcolor=black, citecolor = green!50!blue, urlcolor = blue, filecolor=black, pagebackref=false, hypertexnames=false,  pdfpagelabels ]{hyperref}
\fi

\begin{document}

\maketitle
\CSMsetup

\begin{tcolorbox}[parbox=false,enhanced,breakable]
\section[Abstract]{Abstract}\label{sb:summary}
This article provides an introduction to the theory of mechanism design and its application to engineering problems. Our aim is to provide the fundamental principles of the theory of mechanism design for control engineers and theorists along with the state-of-the-art methods in engineering applications. We start our exposition with a brief overview of game theory highlighting the key notions that are necessary to introduce mechanism design, and then we offer a comprehensive discussion of the principles in mechanism design. Finally, we explore four key applications of mechanism design in engineering, i.e., communication networks, power grids, transportation, and security systems. 
\end{tcolorbox}


\section{Introduction}

Over the last seventy years, the theory of mechanism design was developed as an approach to efficiently align the individuals' and system's interests in problems where individuals have private preferences \cite{Mas_Colell1995}. It can be viewed as the art of designing the rules of a game to achieve a desired outcome. Overall, mechanism design has broad applications spanning many different fields, including microeconomics, social choice theory, computer science \cite{Nisan2007}, and control engineering. Applications in engineering include communication networks \cite{Renou2012}, social networks \cite{Dave2020SocialMedia}, transportation routing \cite{Bian2020}, online advertising \cite{Kakade2013}, smart grid \cite{Samadi2012}, multi-agent systems \cite{Shoham2008}, and resource allocation problems \cite{Hurwicz2006}.

\begin{figure}[ht]
    \centering
    \includegraphics[scale = 0.35]{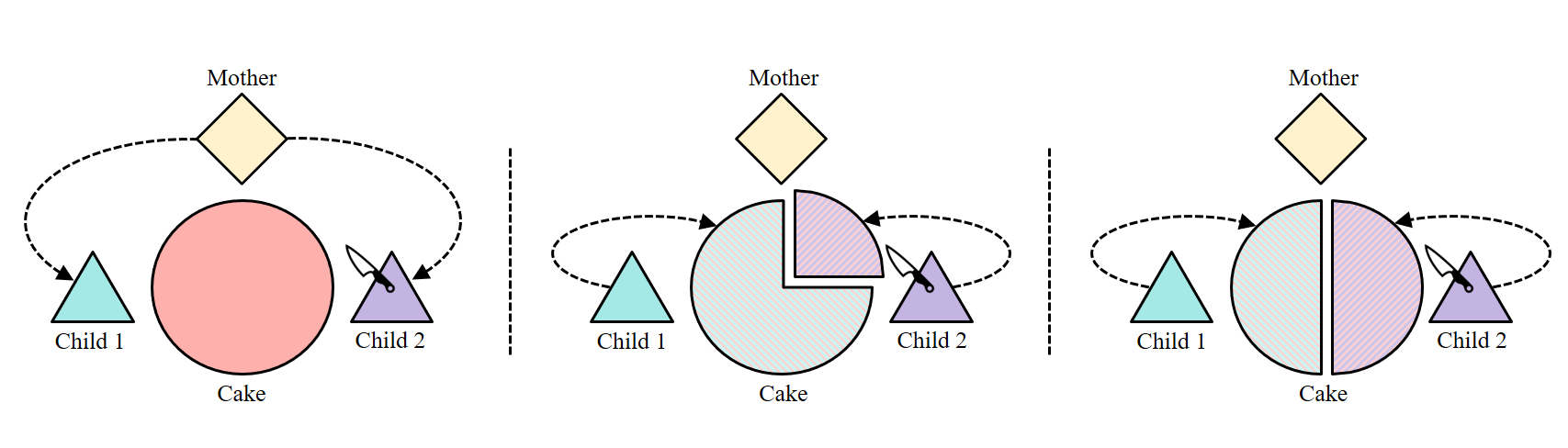}
    \caption{A visualization of the cake example in mechanism design.}
    \label{fig:cake}
\end{figure}

Eric Maskin has provided an example that best illustrates the theory of mechanism design in simple terms \cite{Maskin2019}. Suppose we want to divide a cake between two children, e.g., Mary and Bill (see Fig. \ref{fig:cake}). Our intend is to divide the cake fairly so that each child is satisfied with their portion. Obviously, one way for a fair division is when Mary thinks she has at least half of the cake and so does Bill. However, \emph{how can we achieve such a fair division?} If we are sure that the children see the cake the same way we do, we can just cut it in half and give each child one of the pieces. Mary and Bill will each think that they have half the cake and they will live happily ever after. In reality though, Mary and Bill can be expected to see things differently than us. Children do not always regard our ``fair" division as really fair. Bill might think that Mary's piece is bigger and feel a bit deprived. So, even if we intend to accomplish a fair division, in practice, we are not in the position to obtain it because we know nothing about the children's perspective. Do they see it as we do or not? \emph{Is there a ``mechanism" (e.g., a protocol) that if followed will result in a fair division even if we do not have enough information about the fairness of the division?} Well, one potential mechanism is to  have Bill cut the cake and then Mary would choose the portion she would like. The above mechanism is called \emph{divide-and-choose mechanism}. Since Bill is accountable for the fair division of the cake he will do his best to have it equally cut. He knows that if the pieces are disparate Mary will choose the bigger one. So the cake being equally cut, Mary chooses her portion and she is happy with that as well as Bill who chooses the other one. Hence, through the divide-and-choose mechanism we have accomplish the desired outcome.

The theory of mechanism design represents the confluence of microeconomics \cite{Laffont1996,Clarke1971,Groves1975,Maskin1999,Green1977,Groves1973,Groves1977,Maskin2008,Maskin2000,Reichelstein1988,Armstrong1996,Mathevet2010}, and social choice theory \cite{acemoglu2008political,acemoglu2010dynamic}, while it equally draws from auction \cite{Vickrey1961}, optimization \cite{bandi2012tractable,Bandi2014}, and game theory \cite{Roughgarden2019,Dasgupta1979,Maskin2002,Richter2019,Condorelli2012,Thomson1996}. The theory was developed for the implementation of system-wide optimal solutions to problems involving multiple rational individuals (agents), each with private information about preferences and with conflicting interests \cite{Mas_Colell1995}. It all started by Leonid Hurwicz, who asked what is the best way for a centralized entity who manages a system consisted of rational and intelligent agents to reach a decision when its efficiency depends on the availability (and thus, truthfulness) of the agents' information. Hurwicz's rigorous answer established the theoretical framework to study this problem and other similar ones. The key idea behind his seminal work was to recognize that the solution necessarily depends on the agents' behavior and how they value their information. In economics, this translates to ``a rational and intelligent agent will act as a utility maximizer and will not report their private information truthfully without a guaranteed compensation." Hurwicz developed a methodology to elicit the private information of any agent by offering  appropriate incentives. Hurwicz's theoretical framework was revolutionary both in economics and engineering, and many other scholars started expanding the theory. Rober Myerson and Eric Maskin both contributed immensely into the theory and expanded the mathematics behind ``reverse-engineering" the process of achieving a desirable goal (e.g., social welfare, revenue). The importance, significance, and impact of Hurwicz, Myerson, and Maskin's work was recognized and were awarded the Nobel Memorial Prize in Economic Sciences in 2007 \cite{NobelPrize2007}.


This article has two main objectives: (1) to provide a tutorial of the theory of mechanism design, and (2) to present how the theory of mechanism design can yield solutions to engineering problems. We start our exposition by providing a brief overview on key notions of game theory. Next, we offer the general framework and fundamental principles of mechanism design. Then, we provide mechanism design problem formulations of different engineering applications. Finally, we offer some concluding remarks and a discussion of the future of mechanism design in control engineering.

\section{Game Theory}

Game theory is arguably one of the cornerstones of economics for the study of competition, strategic behavior, and lies at the intersection of mathematics and social science. The first theoretical formulation of a game is thanks to John von Neumann \cite{von_Neumann2007}. Von Neumann's work established the notion that every problem in economics (for that matter in engineering and computer science) is essentially a competition of resources between selfish agents, each striving to make a decision that benefits them only. Thus, von Neumann introduced game theory as a means to study ``interactions in the presence of conflict of interest" \cite{Lam2016}. Game theory models the conflicting (or cooperative) interaction between agents (also referred to as ``players") and provides a principled way of predicting the outcome of this interaction using equilibrium analysis. The study of games is most appealing and intuitive to scientists, engineers, and scholars across many fields as it provides the mathematics to answer what a selfish agent (human, corporation, machine) is going set to decide and how under different yet certain conditions. 

Although game theory is quite extensive, in this article, we focus on providing only a snapshot of the key notions and a simple classification of games (see Fig. \ref{fig:game_theory_visualization}). A \emph{game} is a mathematical model of the strategic interaction of at least two agents (e.g., bidders in an auction, corporations, countries, robots, autonomous cars), whose \emph{actions} (or decisions) can affect the other agent's \emph{payoff}. The agents play against each other, often competing over the utilization of a limited resource (e.g., Tragedy of the Commons \cite{hardin1968tragedy}), by taking an action, which defines the agent's behavior in the game; depending on the application, an action can be either a bid in an auction, or the selection of a route on the road network, or simply what to bet on a coin flip. An agent's action leads to a payoff, meaning that there exists a function that maps the agents' actions to a real number. For example, a vehicle may have to decide on route A or route B. The payoff from choosing either route can be in terms of travel time, and so route A may result to 30 minutes and route B may result to 25 minutes. These actions can be taken \emph{simultaneously} (e.g., rock-paper-scissors), or \emph{sequentially} (e.g., monopoly, chess).
Agents may \emph{cooperate} towards reaching an ideal solution (e.g., signing a contract). This cooperation can take the form of an alliance to find the ``common ground" in terms of actions/strategies. If agents do not choose or are unable to cooperate, then the game is called \emph{non-cooperative} and it most naturally models strictly-competitive ``play" among all agents.

Due to the broad applicability of game theory, there have been a myriad of applications in economics and engineering as well as in computer science, and so specific games have been given a name. For example, \emph{normal-form games} (we offer an overview of these games in \nameref{SIDEBAR:GT_normalformgames}), \emph{matrix games}, \emph{differential, static, dynamic games} \cite{frihauf2011nash,basar1974informational,ratliff2012pricing,kokolakis2021bounded,li2022confluence}, \emph{Bayesian and stochastic games} \cite{fotiadis2022recursive}, \emph{zero-sum games} \cite{bogosyan2022zero,zhai2021data}, and \emph{one-shot and repeated games}, \emph{Stackelberg games} \cite{von2010market,kampezidou2021online}, \emph{finite and continuous games} \cite{ratliff2016characterization}, and \emph{hybrid games} \cite{tomlin2000game,lygeros1995game}.

\begin{figure}[ht]
    \centering
    \includegraphics[scale = 0.4]{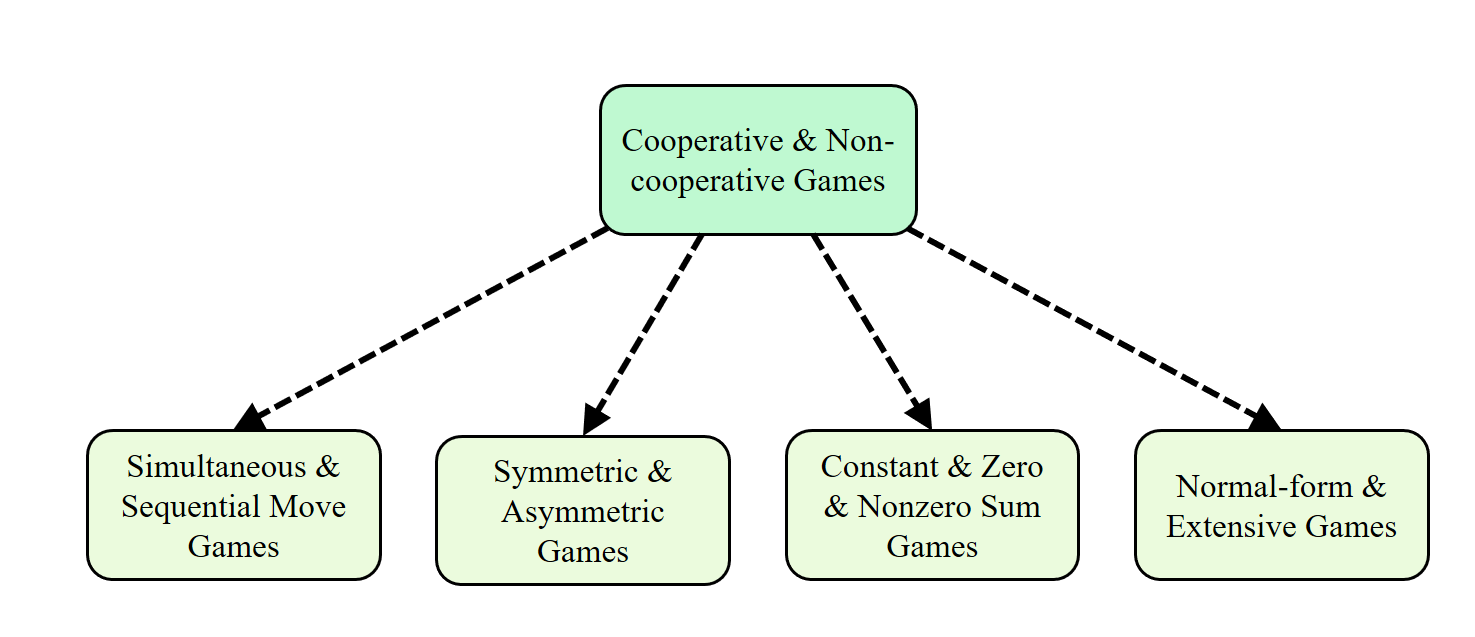}
    \caption{A classification of the different types of games.}
    \label{fig:game_theory_visualization}
\end{figure}

\begin{tcolorbox}[parbox=false,enhanced,breakable]

\sidebars

\section{Sidebar: Games \& the Nash Equilibrium}\label{SIDEBAR:GT_normalformgames}

We present an overview of important notions from non-cooperative game theory.
 A finite normal-form game is a tuple $\mathcal{G} = \langle \mathcal{I}, \mathcal{S}, (u_i)_{i \in \mathcal{I}} \rangle$, where $\mathcal{I} = \{1, 2, \dots, n\}$ is a finite set of $n$ agents (most commonly referred to as players) with $n \geq 2$; $\mathcal{S} = S_1 \times \dots \times S_n$, where $S_i$ is a finite set of actions available to player $i \in \mathcal{I}$ with $s = (s_1, \dots, s_n) \in \mathcal{S}$ being the strategy profile; $u = (u_1, \ldots, u_n)$, where $u_i : \mathcal{S} \to \mathbb{R}$, is a real-valued payoff (or utility) function for player $i \in \mathcal{I}$. 

Games model the \emph{strategic interactions} of competing players. These interactions rely on the information of the players. Thus, in game theory ``who knows what" plays a crucial role. If all players in a game have full access to all available information (payoff or utility functions, players' strategies, and game's dynamics), then we say this is a game with \emph{complete information}. In contrast, even if one agent has limited or missing information, then we say that this is a game with \emph{incomplete information}. If at least one agent does not have access to some information of the game (e.g., strategies of the other players), then we say that this is a game with \emph{imperfect information}. An important notion in game theory is the notion of \textit{common knowledge}, which characterizes a game's information as follows: if every player knows a specific information (e.g., an action), then we can expect any other player also knows that every other player knows it as well.

One of the key assumptions in game theory is that the players are \textit{rational.} A player is said to be rational if they always make decisions in pursuit of their own objectives (e.g., maximize their own expected payoff). Another key assumption in game theory is that the players are \textit{intelligent.} This implies that each player in the game knows everything about the game and they are competent enough to make any inferences about the game.

In all games, players make decisions and reach an \textit{equilibrium.} We call different notions of equilibria a \emph{solution concept}. One such equilibrium is the \emph{dominant strategies equilibrium}, in which any agent has a strategy that regardless of what other players might decide to do, this strategy is the best possible (results in the highest payoff). The dominant strategies equilibrium is quite strong and it can be hard (almost impossible) to have it in a game of competing players under different scenarios. The most celebrated solution concept in game theory is the \emph{Nash equilibrium} (NE). A player's NE strategy is the best response to the NE strategies of the other players. In other words, a player cannot be better off if they depart from their NE strategy if all other players chose their NE strategies. Thus, no player has an incentive to deviate from a NE strategy. More formally, let $S_{i}$ be the set strategies of player $i$,  $s_i, s_i ' \in S_{i}$ be two strategies of player $i$, and $S_{- i}$ be the set of all strategy profiles of the remaining players. Then, $s_i$ strictly dominates $s_i '$ if, for all $s_{- i} \in S_{- i}$, we have $u_i(s_i, s_{- i}) > u_i(s_i ', s_{- i})$. Also, a strategy is (strictly) dominant if it (strictly) dominates any other strategy. A player $i$'s best response to the strategy profile
    \begin{equation}
        s_{- i} = (s_1, \dots, s_{i - 1}, s_{i + 1}, \dots, s_n)
    \end{equation}
is the strategy $s ^ *_i \in S_i$ such that $u_i(s ^ *_i, s_{- i}) \geq u_i(s_i, s_{- i})$ for all $s_i \in S_i$. A strategy profile $s^*=(s^*_1, \dots, s^*_n)$ is a Nash equilibrium (NE) if, for each player $i$, $u_i(s^*_i, s^*_{- i}) \ge u_i(s_i ', s^*_{- i}),$ for all $s_i '\in S_i$.

Next, for completeness, we define the notion of Pareto domination. An outcome of a game is any strategy profile $s \in \mathcal{S}$. Intuitively, an outcome that Pareto dominates some other outcome improves the utility of at least one player without reducing the utility of any other. Let $\mathcal{G}$ and $s ', s \in \mathcal{S}$. Then a strategy profile $s '$ Pareto dominates strategy $s$ if $u_i(s ') \geq u_i(s)$, for all $i\in \mathcal{I}$, and there exists some $j \in \mathcal{I}$ for which $u_j(s ') > u_j(s)$. Pareto domination is a useful notion to describe the social dilemma in a game. However, Pareto-dominated outcomes are often not played in game theory; a NE will always be preferred by rational players. For further discussion of the game theory notions presented above, see \cite{leyton2008essentials,shoham2008multiagent}.

\end{tcolorbox}

\section{The History of Mechanism Design}

Game theory is concerned with the analysis of games. Mechanism design, on the other hand, involves designing games with desirable outcomes. Mechanism design was considered a consequence of debates about the relative merits of socialism, communism, and capitalism, the most important of which was the \emph{socialist calculation debate} \cite{boettke2000socialism}. Mechanism design exhibited an attempt to provide a scientific basis for addressing the above debate by constructing a theoretical framework for considering systems other than capital markets for allocating the means of production \cite{Hurwicz_website}. It also took a mathematically rigorous approach to the comparison of a specific arrangement to capitalism in terms of efficacy and productivity. In the last half-century economics has adopted the study of mechanism design as the systematic analysis of resource allocation in institutions and processes. The above extremely fundamental development reveals the roles of information, communication, control, incentives and agent processing capacity in decentralized resource allocation. Moreover, it allows the identification of sources of market failure. Leo Hurwicz, a Polish-American economist and mathematician, was the first who introduced the concept of \emph{incentive compatibility} and then provided a methodology for mechanisms that are incentive compatible and how they can guarantee the desired outcomes \cite{Hurwicz1960}. Hurwicz's contributions in establishing mechanism design's theoretical foundations were key in providing efficient solutions to resource allocation problems. Jean-Jacques Laffont, a French economist, through his studies in public and information economics, participated in the translation of the foundational economic theory into the language and tools which today appear not only in game theory but also in studies of the organization of firms and markets as well as in the applied economics of regulation, taxation, and public goods provision \cite{Laffont2009}. Pure and applied research in economics was connected through the studies about transactions among economic agents in terms of information and incentives. The discipline of economics itself has been transformed including observers and commentators on economic systems as well as architects who design incentives and engineers who implement and test institutions. Other excellent historical surveys on the theory of mechanism design are reported in \cite{mcfadden2009human,narahari2014game}.

\begin{figure}[ht]\setcounter{figure}{2}
    \centering
        \includegraphics[width = 1 \columnwidth]{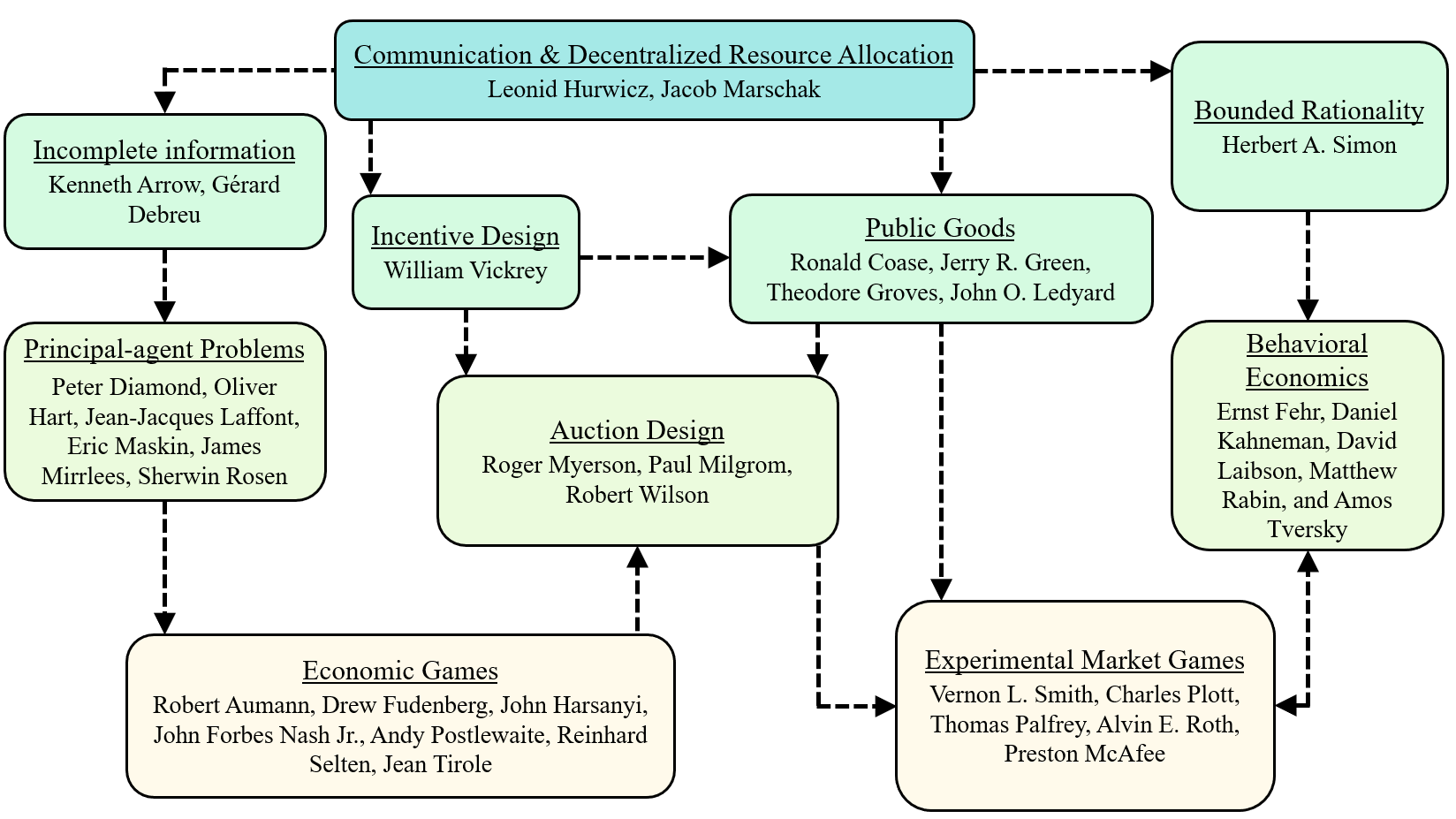}
        \caption{A snapshot of the key theoretical ``chapters" of the theory of mechanism design and its related fields.}
        \label{fig:mechanismdesign_outline}
\end{figure}

The theory of mechanism design since the 1950s has been rigorously studied by numerous economists and mathematicians to provide insights and solutions to different economic topics (e.g., public goods, markets, auctions). The theory started with the seminal contributions of Leonid Hurwicz and Jacob Marschak (see Fig. \ref{fig:mechanismdesign_outline}), who both were interested in resource allocation problems (e.g., communication) and the means of controlling (through incentives) agents. In parallel, Kenneth Arrow, G\'{e}rard Debreu, and Herbert A. Simon also worked on problems with incomplete information and how to bound rationality. 

In 1961, William Vickrey's seminal work on auctions was published paving the way for Hurwicz's theoretical framework to be applied as a means in designing incentives based on the agents' information for a simple yet formidable problem of an auction. Both Arrow and Debreu's work was instrumental in establishing the interconnected relation of information and decision-making, its role in influencing behavior in the efficient allocation of limited resources. 

Much later in the 1970s and 1980s, Peter Diamond, Oliver Hart, Jean-Jacques Laffont, Eric Maskin, James Mirrlees, and Sherwin Rosen worked independently on principal-agent problems focusing on how one can design a contract between a principal (e.g., institution, corporation) and a rational agent efficiently. As a continuation of Vickrey's work, Ronald Coase, Jerry R. Green, Theodore Groves, and John Ledyard made significant contributions in the design of incentives for public goods problems (e.g., road infrastructure, public parks, television and radio broadcasts, national security). Furthermore, Roger Myerson, Paul Milgrom, and Robert Wilson expanded the Vickrey auction to address more complicated scenarios and complex problems.

At the same time, many economists continued to develop and study games; a few examples are Robert Aumann, Drew Fudenberg, John Harsanyi, John Forbes Nash Jr., Andy Postlewaite, Reinhard Selten, and Jean Tirole. Game theory and mechanism design address the interactions among strategic agents. An important extension of both theories is the development of methodologies for agents who might not act as rational and intelligent agents. Key questions in this area are: (i) Do agents make optimal decisions at all times? (ii) Do agents make sacrifices when deciding for their utility? (iii) Do agents always optimize for their own benefit? To answer these questions, the fields of behavioral economics and experimental game theory were developed, first by Herbert A. Simon, and then by notable contributors as Ernst Fehr, Daniel Kahneman, David Laibson, Matthew Rabin, and Amos Tversky, as well as Vernon L. Smith, Charles Plott, Thomas Palfrey, Alvin E. Roth, and Preston McAfee. Several survey papers of the theory of mechanism design can be found in \cite{Klemperer2004,Krishna2002,Maskin2004,Milgrom2004}.

Although we have only focused on the history of mechanism design from the economics point of view, there have been numerous key contributions in engineering and computer science. Thus, we have dedicated the second part of this article to stress the contributions and engineering applications of mechanism design. 

\section{The Theory of Mechanism Design}

Most generic control systems can be viewed as a specification of how decisions are made as a function of the information that is known by the agents in the system \cite{Myerson1981,Thomas2002,Johari2009,Johari2004,Kelly1998}. What interests us in most cases is \emph{efficiency}, i.e., realizing the best possible allocation of resources with the best use of information to achieve an outcome where collectively agents are satisfied, and there is no overutilization of the system's resources \cite{Maskin2002}. One key challenge in ensuring efficiency in a control system is the fact that different agents may have conflicting interests and act selfishly. In other words, systems that incorporate human decision-making, if remained uninfluenced, are not guaranteed to exhibit optimal performance. This is well-known to be the case in control theory, and economics \cite{Myerson2013,Brown2018}. There are various different theories and approaches that attempt to guarantee efficiency in such systems and can provide solutions of varying degrees of success. One way to study such problems is \emph{information design} (see \nameref{SIDEBAR:MD_informationdesign}). Another theory is mechanism design, in which we are concerned with how to implement system-wide optimal solutions to problems involving multiple selfish agents, each with private information about their preferences \cite{Nisan2001,Diamantaras2009}. For example, within the context of mobility, agents are the travelers, and their private information can be either tolerance to traffic delays, value of time, preferred travel time, or any disposition to a specific mode of transportation \cite{zhao2019enhanced}. Given that each traveler/driver/passenger ``competes'' with everyone else to reach their destination first, we want to ensure that given this inherent conflict of interest, we can still guarantee uncongested roads, no traffic accidents, and no travel time delays. Mechanism design can help us design the rules of systems where information is decentralized (different agents know different aspects of the system), and agents do not necessarily have an immediate incentive to cooperate \cite{Borgers2015}. In particular, mechanism design helps us design rules that align all agents' decision-making by providing the right incentives to achieve a well-defined objective for the system (e.g., aggregate optimal performance, system-level efficiency). Thus, mechanism design entails solving an optimization problem with sometimes unverifiable and always incomplete information structure \cite{Maskin2008}. We call such a problem \emph{an incentive design and preference elicitation problem}.

\begin{tcolorbox}[parbox=false,enhanced,breakable]

\sidebars

\section{Sidebar: Information Design}\label{SIDEBAR:MD_informationdesign}

In this sidebar, we offer a quick overview of an alternative approach to mechanism design based on the work reported in \cite{S:Bergemann2019,S:Taneva2019}. As we have seen so far, information plays a crucial role in game-theoretic models and in mechanism design. However, information in such cases is used to evaluate the utility functions and study equilibria. In mechanism design, in particular, information is used to design the best-possible payment functions to guide efficient allocations. In contrast, in \emph{information design}, the social planner manages a system's information (instead of its resources) and it is their task to devise a careful process for the efficient allocation of information (instead of monetary incentives for the allocation of resources). Thus, the goal of information design is to influence the agents' behavior based on how much information may become available. This draws key insights from behavioral economics as it treats the agents' cognitive abilities to adapt their behavior based on what is available to them.

It is standard in mechanism design to consider as a given the ``informational environment" (e.g., who knows what and who does not know what yet seeks to learn), and focus in the design of appropriate incentives based on the desirable outcome (equilibrium behavior) among strategic agents. Information design simply considers non-fixed informational environments and sets the set of rules in which the agents and the social planner commit to respect. Of particular interest are environments with incomplete information as strategic compete over resources against each other and seek to learn more about their competitors (other agents). For example, an agent can improve their strategy if they have more accurate beliefs about the payoffs and states of other agents.

The last ten years, information design has been growing rapidly finding key applications in economics, engineering, and finance. One example is the study of optimal design of information under a Bayesian framework between two agents who attempt to communicate over a network \cite{S:Kamenica2011,S:Brocas2007,S:Rayo2010}. Other applications are grade disclosure and matching markets \cite{S:Ostrovsky2010}, voter mobilization \cite{S:Alonso2016}, traffic routing \cite{S:Das2017}, rating systems \cite{S:Duffie2017}, and transparency regulation \cite{S:Asquith2013} in financial markets, price discrimination \cite{S:Bergemann2015}, and stress tests in banking regulation \cite{S:Inostroza2017}.

\end{tcolorbox}

\subsection{The Building Blocks of Mechanism Design}

We start our exposition by considering a system consisting of a finite group of agents, each competing with each other for a limited, fixed allocation of resources. Each agent evaluates different allocations based on some private information that is known only to them. We consider a \emph{social planner}, playing the role of a centralized entity, whose task is to align the selfish and conflicting interests of the agents with the overall system's objective (e.g., an efficient allocation of resources or the maximization of social welfare). As illustrated in Fig. \ref{fig:mechanism_design_visualization}, there are four components. There is a group of agents, each making a decision based on their personal information. Decisions then are reported as messages to the social planner who is tasked to design the rules of which it can be determined what each agent gets.

\begin{figure}[ht]\setcounter{figure}{3}
    \centering
    \includegraphics[scale = 0.35]{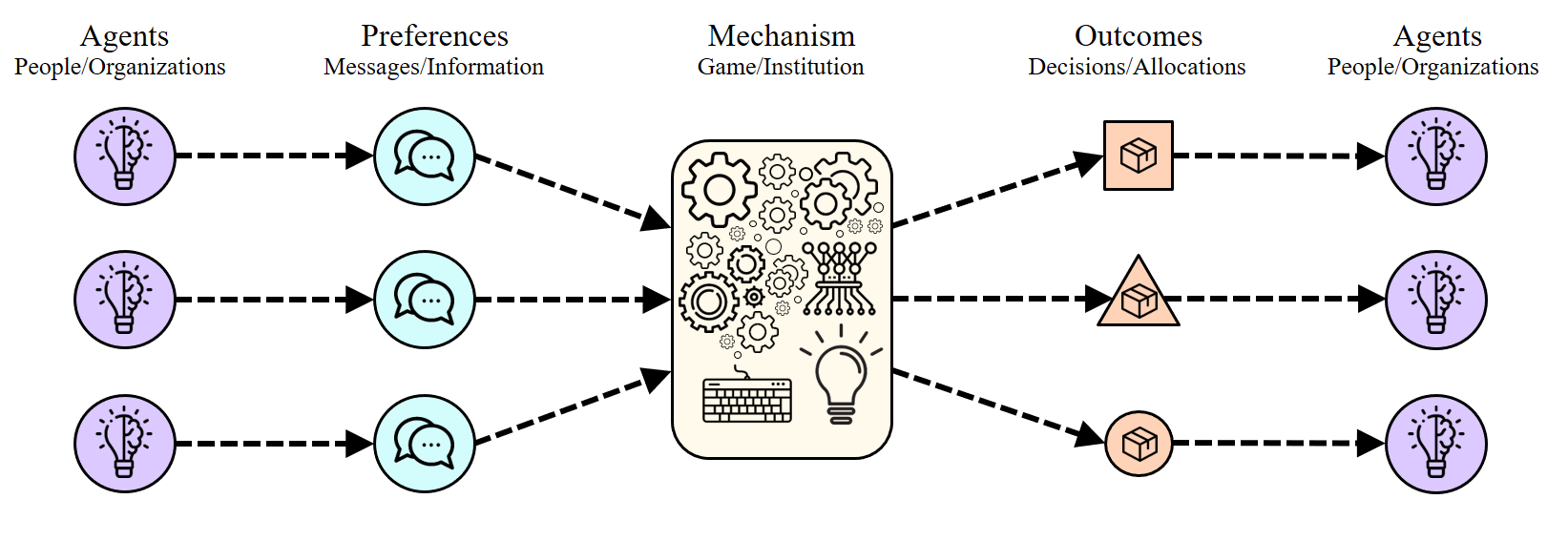}
    \caption{A visualization of how an arbitrary control system (agents, preferences, allocations) can be viewed under a mechanism design framework.}
    \label{fig:mechanism_design_visualization}
\end{figure}


Next, we provide a formal mathematically presentation of the social planner's task through the lens of optimization theory.
We consider a set of selfish agents $\mathcal{I}$, $|\mathcal{I}| = n \in \mathbb{N},$ with preferences over different outcomes in a set $\mathcal{O}$. Each agent $i \in \mathcal{I}$ is assumed to possess private information, denoted by $\theta_i \in \Theta_i$. Since an agent $i$'s $\theta_i$ can influence their decision-making in a significant way, we call $\theta_i$ the \emph{type} of agent $i$. We write $(\theta_i)_{i \in \mathcal{I}} = \theta,$  $\theta\in \Theta$, where $\Theta = \prod_{i \in \mathcal{I}} \Theta_i,$ to represent the type profile of all agents. An agent $i$'s preferences over different outcomes can be represented by a utility function $u_i : \mathcal{O} \times \Theta_i \to \mathbb{R}$. Although the exact form of $u_i$ can vary depending on the application of the problem \cite{Tavafoghi2016,Tang2011,Bitar2016,Kazumura2020}, what is common in the literature \cite{Shoham2008,Borgers2015,Nisan2007} is a \emph{quasilinear function} of the form
    \begin{equation}\label{eqn:defn_utility}
        u_i(o, \theta_i) = v_i(o, \theta_i) - p_i,
    \end{equation}
where $v_i : \mathcal{O} \times \Theta_i \to \mathbb{R}_{\geq 0}$ represents an arbitrary valuation function, and $p_i \mapsto \mathbb{R}$ is a monotonically increasing function. If outcome $o \in \mathcal{O}$ represents an allocation of a resource, then $p_i$ can be thought of as a transfer of agent $i$'s wealth or a cost imposed to agent $i$ for that particular allocation $o$. Intuitively, a quasilinear function defined as in \eqref{eqn:defn_utility} ensures that the marginal value of $v_i$ does not depend on how large $p_i$ becomes, and vice-versa. Furthermore, \eqref{eqn:defn_utility} assumes $u_i$ is linear with respect to $p_i$. Next, we can naturally define the \emph{social welfare} as the collective summation of all agents' valuations, i.e.,
    \begin{equation}\label{eqn:social_welfare}
        w(o, \theta) = \sum_{i \in \mathcal{I}} v_i(o, \theta_i).
    \end{equation}
If our system objective is to maximize $w$, then immediately we observe that there is an important obstacle, i.e., any agent $i$ may misreport their type $\theta_i$ in the hopes to increase their own utility. So, the question is now: How can we incentivize agents to truthfully report their type? The answer is through the appropriate design of $p_i$. Next, we outline the building blocks that can help us design $p_i$. Formally, we can define a \emph{mechanism} as the tuple $\langle f, p \rangle$ composed of a \emph{social choice function} (SCF) $f : \Theta \to \mathcal{O}$ and a vector of \emph{payment functions} $p = (p_i)_{i \in \mathcal{I}}$, with $p_i : \Theta \to \mathbb{R}$. In words, a mechanism $\langle f, p \rangle$ defines the rules of which we can implement a system objective by mapping the agents' types to an outcome (in the means of the SCF) while using the payments to ensure the optimality or efficiency of that outcome. It is important to note here that the above is an example of \emph{direct mechanism}, in which information is directly communicated between the agents and the social planner (see \nameref{SIDEBAR:fundamental} for other specifications). Next, we state the social planner's problem as follows
    \begin{gather}
        \max_{o \in \mathcal{O}} w(o, \theta) \label{eqn:obj_function} \\
        \text{subject to: } \hat{\theta}_i = \theta_i, \quad \forall i \in \mathcal{I}, \label{eqn:constraint_truth} \\
        \sum_{i \in \mathcal{I}} v_i(o, \theta_i) \geq \sum_{i \in \mathcal{I}} v_i(o ', \theta_i), \quad \forall o ' \in \mathcal{O}, \label{eqn:constraint_efficiency} \\
        \sum_{i \in \mathcal{I}} p_i(s(\theta)) \geq 0, \quad \forall \theta \in \Theta, \label{eqn:constraint_budget} \\
        v_i(f(s(\theta))) - p_i(s(\theta)) \geq 0, \quad \forall i \in \mathcal{I}, \quad \forall \theta \in \Theta, \label{eqn:constraint_IR}
    \end{gather}
where $\hat{\theta}_i$ denotes the reported type of agent $i$, $s(\cdot)$ is the equilibrium strategy profile (e.g., Nash equilibrium). Constraints \eqref{eqn:constraint_truth} ensure the truthfulness in the agents' reported types, \eqref{eqn:constraint_efficiency} imposes an efficiency condition, \eqref{eqn:constraint_budget} makes certain that no external payments are required, and \eqref{eqn:constraint_IR} incentivizes all agents to voluntarily participate in the mechanism. If we could know for certain the true types of all agents, then we would be able solve the optimization problem \eqref{eqn:obj_function} - \eqref{eqn:constraint_IR} using standard optimization techniques. However, as this is unreasonable to expect from selfish agents, the social planner needs to elicit $\theta = (\theta_i)_{i \in \mathcal{I}}$ by designing the appropriate $p = (p_i)_{i \in \mathcal{I}}$.

The social planner faces now two critical questions: the \emph{preference aggregation}, which asks what is the best outcome $o \in \mathcal{O}$ for any given type profile $\theta \in \Theta$, and the \emph{information elicitation}, which asks how one can extract truthfully the type of any agent $\theta_i \in \Theta$. The theory of mechanism design essentially helps us answer both questions by providing the mathematical framework to construct mechanisms $\langle f, p \rangle$ that can achieve our desirable outcome. In the next subsection, we discuss one such mechanism that elicits the private information of agents truthfully.

\begin{tcolorbox}[parbox=false,enhanced,breakable]

\sidebars

\section{Sidebar: Fundamental Results}\label{SIDEBAR:fundamental}

One key characteristic of mechanism design is the communication of information in the system. Agents have private information, which is vital to the social planner's objective. Part of any mechanism is to specify how the private information is communicated from the agents to the social planner, and thus all mechanisms fall under two categories: \emph{direct} and \emph{indirect}. Given any system, if the agents report their private information (preferences) directly to the social planner, then we say that the agents' preferences are observable to the social planner. In contrast, if the agents do not or cannot report their private information to the social planner, then the social planner has to ``observe" the agents' preferences indirectly through signals or behavior. Formally, an indirect mechanism is defined as the specification of $\langle \mathcal{M}, g \rangle$, a collection of messages $\mathcal{M} = (\mathcal{M}_i)_{i \in \mathcal{I}}$ and an outcome function $g$. A direct mechanism is defined as the tuple $\langle f, p \rangle$. One key question in mechanism design now is the following: ``If an outcome can be implemented in an indirect mechanism, then can it also be implementable in a direct mechanism where information (types) is observable?" This is answered by the \emph{revelation principle}.

The revelation principle is one the most fundamental and significant results in the theory of mechanism design. It serves as the cornerstone establishing that the solution of any indirect mechanism can most surely be replicated by a direct mechanism. This allows us to limit the scope of how many mechanisms we need to investigate and focus rather on mechanisms in which agents communicate privately and directly with the social planner. Thus, the goal remains to elicit the private information truthfully \cite{S:Dasgupta1979,S:gibbard1973,S:harris1981,S:myerson1979incentive,S:myerson1982optimal}. Mathematically, we have the following theoretical setup: suppose some arbitrary mechanism $\langle \mathcal{M}, g \rangle$ implements some SCF $f$ in a \emph{dominant strategy equilibrium}. This means that SCF $f$ is truthfully implementable in dominant strategy equilibrium if and only if for each agent $i \in \mathcal{I}$, there exists functions $m_i : \Theta_i \to \mathcal{M}_i$ such that for $\theta_i \in \Theta_i$ and $g(m(\theta)) = f(\theta)$ the profile $(m_i(\theta_i))_{i \in \mathcal{I}}$ is a dominant strategy. Hence, it can be seen that the revelation principle holds immediately after noting that $f(\theta) = g(m(\theta))$ for each $\theta \in \Theta$. There are two key theoretical insights we can draw from the revelation principle: (i) for the implementation of SCF it is only sufficient to focus to a system's main attributes; (ii) in systems, decentralization cannot prevail centralization. As we will see in later sections though, decentralization and the specifics of a problem can offer useful insights and are worth to investigate indirect mechanisms.

\end{tcolorbox}

\subsection{The Vickrey-Clarke-Groves Mechanism}

A well-established and broadly-used mechanism that has been successful in widely different applications (e.g., auctions, public projects, and cost-minimization problems) is the Vickrey-Clarke-Groves (VCG) mechanism \cite{Vickrey1961,Clarke1971,Groves1973}. The VCG mechanism ensures the existence and implementation of a dominant strategy equilibrium, which is an efficient solution and allows selfish agents to make a decision (alternatively choose a strategy) that is best no matter what other agents may decide. Agents are also incentivized to truthfully report their private preferences and have no reason (e.g., chance of receiving negative utility) not to participate in the mechanism. However, the VCG mechanism is known to be an extravagant mechanism, i.e., it can generate big surpluses.

In the previous subsection, we reviewed the main concepts of mechanism design and formulated the incentive design and preference elicitation problem. In words, we asked ``How can we design the payments $p = (p_i)_{i \in \mathcal{I}}$ so that every agent makes the decision that agrees with what \emph{we} have chosen as the system's objective (e.g., efficiency)? To answer this question, in this subsection, we review the Vickrey-Clarke-Groves (VCG) mechanism \cite{Vickrey1961,Clarke1971,Groves1973}, one of the most successful mechanisms as it incentivizes agents to be truthful and guarantees efficiency.

As we discussed earlier, a mechanism is a tuple $\langle f, p \rangle$. In a VCG mechanism, the SCF $f$ is defined as an allocation rule (who gets what) based on the optimization problem \eqref{eqn:obj_function} - \eqref{eqn:constraint_IR}, i.e.,
\setcounter{equation}{8}
    \begin{equation}\label{eqn:scf2}
        f(\hat{\theta}) = \arg \max_{o \in \mathcal{O}} W(o, \hat{\theta}_i),
    \end{equation}
where $\hat{\theta} = (\hat{\theta}_i)_{i \in \mathcal{I}}$. In words, assuming that the agents disclose their true information, \eqref{eqn:scf2} provides the social planner, who attempts to maximize the social welfare, with a formal way to compute the allocations of each agent. At the same time, the VCG mechanism charges each agent for their allocation as follows
    \begin{equation}\label{eqn:payments}
        p_i(\hat{\theta}) = \sum_{j \neq i} v_j (f(\hat{\theta}_{- i})) - \sum_{j \neq i} v_j (f(\hat{\theta})),
    \end{equation}
where $\hat{\theta}_{- i}$ denotes the type profile of all agents except agent $i$. Note that the payments defined in \eqref{eqn:payments} do not depend on an agent $i$'s own declaration $\hat{\theta}_i$. Let us assume for a moment that all agents declare their types truthfully. Then, the first sum in \eqref{eqn:payments} computes the value of the social welfare with agent $i$ not participating in the mechanism. The second sum in \eqref{eqn:payments} computes the value of the social welfare of all other agents $j \neq i$ with agent $i$ participating in the mechanism. Thus, agent $i$ when they report $\hat{\theta}_i$ are made to pay the \emph{marginal effect} of their decision (in our case that is agent $i$'s reported type $\hat{\theta}_i$). In other words, this particular design of the payments in \eqref{eqn:payments} internalizes an agent $i$'s social externality, i.e., agent $i$'s impact on every other agents' welfare.

The VCG mechanism represented by the SCF $f$ defined by \eqref{eqn:scf2} and the payment functions $p$ defined by \eqref{eqn:payments} satisfies the following properties:
    \begin{enumerate}
        \item For any agent, truth-telling is a strategy that dominates any other strategy that is available for that agent. We say then that truth-telling is a \emph{dominant strategy}. Note that such strategies are ``always optimal'' no matter what the other agents decide.
        \item The VCG mechanism successfully aligns the agents' individual interests with the system's objective. In our case, that objective was to maximize the social welfare of all agents. We call this property, \emph{economic efficiency}.
        \item For any agent, the VCG mechanism incentivizes them to voluntarily participate in the mechanism as no agent loses by participation (in terms of utility).
        \item The VCG mechanism ensures no positive transfers are made from the social planner to the agents. Thus, the mechanism does not incur a loss. We call this \emph{weakly budget balanced}.
    \end{enumerate}
The VCG mechanism essentially ensures the realization of a \emph{socially-efficient outcome}, i.e., satisfying Properties (1) - (3), in a system of selfish agents, where each possesses private information.  The VCG mechanism induces a dominant strategy equilibrium maximizing the social welfare while also making sure no agent is hurt by participating.

\begin{figure}[ht]\setcounter{figure}{4}
    \centering
    \includegraphics[scale = 0.35]{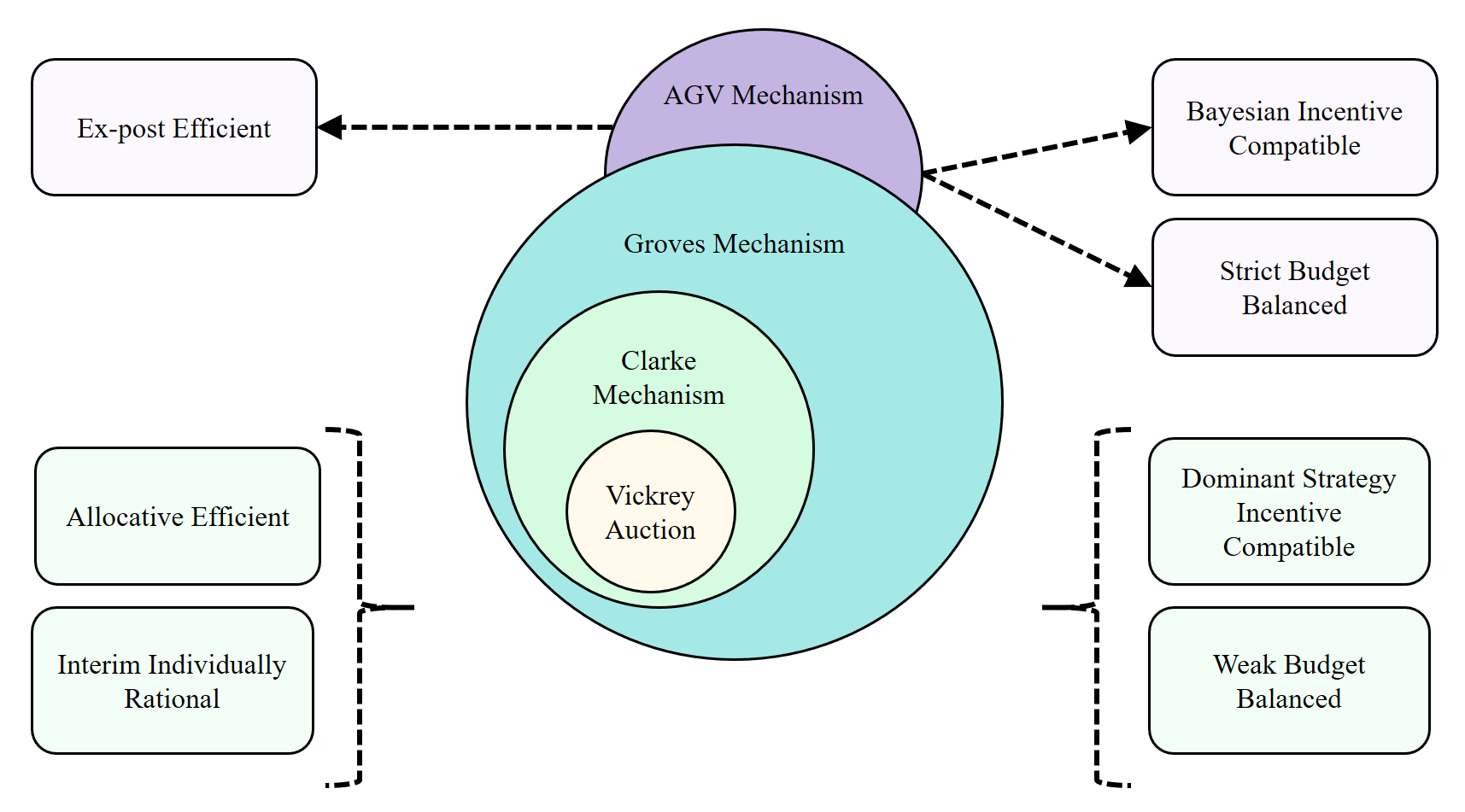}
    \caption{A visualization of the VCG and AGV mechanisms and their possible properties.}
    \label{fig:mechanism_design_properties}
\end{figure}

Other mechanisms exist, each with different properties and tradeoffs between efficiency, optimality, and information. One classic example is the Arrow and d'Aspremont and Gerard-Varet (AGV) mechanism that offers an efficient solution, incentivizes all agents to report their private information truthfully, and most importantly all transactions between the agents and the social planner equal to zero (i.e., budget-balanced). The key characteristic of the AGV is its solution concept \emph{Bayesian-Nash equilibrium} that operates under the assumption of a \emph{common prior}, i.e., agents hold beliefs on what other agents might do. Furthermore, implementation theory focuses on decentralized mechanisms and setting up more rigorously the messaging process between agents and the social planner. Inspired from economics and engineering applications, \emph{distributed mechanisms} (see \cite{Heydaribeni2020,Sinha2014,Sinha2017_a,Sinha2017_CDC}) and \emph{dynamic mechanisms} (see \cite{kotsalis2010robust,schutte2019dynamic,kotsalis2013dynamic}) have also been developed to tackle how decisions can be made locally within a system and also by relaxing some of the strong assumptions on information and prior beliefs, respectively.

We conclude this section with the following remark. Although the main motivation of mechanism design is the microeconomic study of institutions and relies heavily on game-theoretic techniques, it is a powerful theory providing a systematic methodology in the design of systems of asymmetric information \cite{Malikopoulos2021,Malikopoulos2022a}, consisted of strategic agents, whose performance must attain a specified system objective. The remaining article shall present how we can use this theory to design a socially-efficient systems consisting of rational and intelligent agents who compete with each other for the utilization of a limited number of resources.

\section{Engineering Applications}

In this section, we present some engineering applications that mechanism design has been used.

\subsection{Communication Networks}

Communication networks are typically modeled as resource allocation problems with a finite set of strategic agents (e.g., the bandwidth in wired/wireless communication is the resource and a ``network manager" needs to allocate efficiently among the agents). There are many different focused areas in communications such as flow control, routing, channel scheduling or power control. There are numerous ways to model the utility of an agent in a communication network, but one key assumption is to consider network utilities for the agents as a concave function. Naturally, game-theoretic studies  \cite{han2012game,han2019game,saad2009coalitional,saad2012game,alpcan2002cdma,zhu2015game} have been extensively focused on routing/congestion games, networked games, and dynamic games. Main results are the existence and uniqueness of a NE, its computation as well as deriving algorithms to learn/attain a NE. A completely different approach is to formulate the communication network as a \emph{convex optimization problem} (e.g., maximize network utility of all agents) subject to network constraints \cite{Kakhbod2012_TAC,Farhadi2019}. Typically, a communication problem is formulated as follows. We consider a communication network of $n$ strategic agents, where the set of agents is given by $\mathcal{I}$. Each agent is rational and intelligent and possess private information. In addition, each agent $i \in \mathcal{I}$ is endowed with a utility function $u_i : \mathcal{X} \times \mathbb{R}$, where $\mathcal{X}$ is the set of allocation. In such problems, we consider \emph{quasilinear} utility functions of the form
    \begin{equation}
        u_i(\mathbf{x}_i, p_i) = v_i(\mathbf{x}_i) - p_i,
    \end{equation}
where $\mathbf{x}_i = (x_i ^ 1, x_i ^ 2, \dots, x_i ^ {k_i})$ is a $k_i$-dimensional vector and denotes the allocation made to agent $i \in \mathcal{I}$. Intuitively, a quasilinear function defined as above ensures that the marginal value of $v_i$ does not depend on how large $p_i$ becomes, and vice-versa. Furthermore, $u_i$ is linear with respect to $p_i$, which represents the tax paid by agent $i \in \mathcal{I}$. Based on this information so far, we get the following optimization problem
    \begin{gather}
        \max_{\mathbf{x}} \sum_{i \in \mathcal{I}} v_i(\mathbf{x}_i) \\
        \text{subject to: } \sum_{i \in \mathcal{S}_j} h_{i j} (\mathbf{x}_i) \leq c_j, \quad \forall j = 1, 2, \dots, m_c, \\
        \mathbf{x}_i \geq 0, \quad \forall i \in \mathcal{I},
    \end{gather}
where $m_c \in \mathbb{N}$ is the number of network constraints (e.g., capacity of a link in the network) and $\mathcal{S}_j$ is the set of agents associated with the $j$-th constraint (e.g., number of agents using a particular link in the network). For each $j = 1, \dots, m_c$, $h_{i j}$ is a general function that depending on the specifics of the problem may model how we can measure the allocation of bandwidth in the network. Next, agents hold private information that is not known to the social planner. An example is the valuation functions $(v_i)_{i \in \mathcal{I}}$. Thus, the social planner cannot directly address the above maximization problem without the agents' valuation functions. We say information is decentralized among the agents in the communication network and it is the task of the social planner to devise a mechanism to elicit the necessary information from the agents and ensure the efficient allocation of the resources. This is a clear and natural application of \emph{implementation theory} in an engineering problem (see \nameref{SIDEBAR:MD_implementation}).

\begin{tcolorbox}[parbox=false,enhanced,breakable]

\sidebars

\section{Sidebar: Implementation Theory}\label{SIDEBAR:MD_implementation}

In this sidebar, we present the fundamentals of \emph{implementation} theory following the formulation of an indirect and decentralized resource allocation mechanism closely inspired by the framework presented in \cite{S:Hurwicz2006}. One key characteristic of implementation theory is that it considers systems which are informationally decentralized. Thus, the goal is to devise a mechanism that handles this ``asymmetry" in information and provide a set of rules that induce a game. This game will have an equilibrium achieving our desired outcome among strategic agents. So, what are the building blocks of this theory? First, we need to specify a set of messages that all agents have access and are able to use in order to communicate information. Based on this information, agents make decisions which affect the reaction of the network manager. Once the communication between the network manager and the agents is complete, we say that the mechanism induces a game; strategic agents then compete for the network's resources. In this line of reasoning, we define formally below what we mean by indirect mechanism and induced game. An indirect mechanism can be described as a tuple of two components, namely $\langle \mathcal{M}, g \rangle$. We write $\mathcal{M} = (\mathcal{M}_1, \mathcal{M}_2, \dots, \mathcal{M}_n)$, where $\mathcal{M}_i$ defines the set of possible messages of agent $i \in \mathcal{I}$. Thus, the agents' complete message space is $\mathscr{M} = \mathcal{M}_1 \times \dots \times \mathcal{M}_n$. The component $g$ is the outcome function defined by $g : \mathscr{M} \to \mathcal{O}$ which maps each message profile to the output space
    \begin{equation}\setcounter{equation}{2}
        \mathcal{O} = \{(x_1, \dots, x_n), (t_1, \dots, t_n) \; | \; x_i \in \mathbb{R}_{\geq 0}, \; p_i \in \mathbb{R}\},
    \end{equation}
i.e., the set of all possible allocations to the agents and the monetary payments made or received by the agents. The outcome function $g$ determines the outcome, namely $g(\mu)$ for any given message profile $\mu = (m_1, \dots, m_n) \in \mathcal{M}$. The payment function $p_i : \mathcal{M} \to \mathbb{R}$ determines the monetary payment made or received by an agent $i \in \mathcal{I}$. A mechanism $\langle \mathcal{M}, g \rangle$ together with the utility functions $(u_i)_{i \in \mathcal{I}}$ induce a game $\langle \mathcal{M}, g, (u_i)_{i \in \mathcal{I}} \rangle$, where each utility $u_i$ is evaluated at $g(\mu)$ for each agent $i \in \mathcal{I}$. Next, we introduce the notation $m_{- i}$, i.e., the message profile of all agents except agent $i \in \mathcal{I}$, i.e., $m_{- i} = (m_1, \dots, m_{i - 1}, m_{i + 1}, \dots, m_n)$. Next, consider a game $\langle \mathcal{M}, g, (u_i)_{i \in \mathcal{I}} \rangle$. The solution concept of NE is a message profile $\mu ^ *$ such that $u_i (g(m_i ^ *, m_{- i} ^ *)) \geq u_i(g(m_i, m_{- i} ^ *))$, for all $m_i \in \mathcal{M}_i$ and for each $i \in \mathcal{I}$, where $m_{- i} = (m_1, \dots, m_{i - 1}, m_{i + 1}, \dots, m_n)$. Note that a NE requires complete information. But, we can interpret a NE as the fixed point of an iterative process in an incomplete information setting \cite{S:Reichelstein1988,S:Groves1977}. This is in accordance with John Nash's interpretation of a NE, i.e., the complete information NE can be a possible equilibrium of an iterative learning process.

\end{tcolorbox}

Using implementation theory and taking advantage of an optimization-based formulation the focus is to carefully design the outcome function $g$, i.e., design the payment function $p_i$ such that the allocations $\mathbf{x}$ are efficient for all agents. The standard analysis of such problems is as follows: first, we show that at least one NE exists for the game induced by the mechanism $\langle \mathcal{M}, g \rangle$. Then, since agents are considered to be strategic, we need to ensure that the outcome function (and thus, the payment functions) induce \emph{voluntary participation}, i.e., at any NE, no agent may lose utility from participating in the mechanism. So, in other words, all agents may at the very least have neutral utility ($u_i = 0$). Naturally, in communication networks as well as any other engineering applications we need to be mindful of all monetary transactions in the system. Thus, by checking the sum of all the payments of all the agents, we need to ensure it is zero at NE, which in that case we say the mechanism is \emph{budget balanced}. The next two properties are related to the optimal solution of the optimization problem. Can we ensure that all NE of the induced game $\langle \mathcal{M}, g, (u_i)_{i \in \mathcal{I}} \rangle$ are equivalent to the optimal solution? If yes, then we say that the mechanism fully implements the efficient allocation vector $\mathbf{x} ^ *$ at a NE. Rarely, we might have an ideal mechanism that results a unique NE. Finally, depending on the network's topology and other physical constraints of our problem of interest, we need to check feasibility, i.e., the allocation $\mathbf{x} = (x_1, \dots, x_n)$ for each agent is feasible satisfying all the constraints at NE. This methodology has been widely used in the communication networks literature, and so, in Table \ref{tab:Comm_table} we provide a snapshot of recent work that mostly focuses on the utility maximization problem.

In the next subsection, we focus one particularly interesting method that has been studied extensively from economics in communication networks. Of course, other mechanisms have also been investigated in the literature \cite{loiseau2013incentive,shakkottai2008price,ozdaglar2007incentives,menache2011game,acemoglu2022too}.

\begin{table}[ht]
    \centering
    \caption{A summary of recent papers on utility maximization problems in communication networks.}
\begin{tabular}{ |p{1.6cm}|p{4.5cm}|p{2.5cm}|p{2.6cm}|p{1.7cm}|}
\hline
    \multicolumn{5}{|c|}{Communication Networks with a Maximization Problem} \\
\hline
    Reference & Framework & Constraints & Implementation & Budget Balance \\ 
\hline
\cite{Yang2007} & Flow Control & System-wide & Partial & No \\ 
\cite{Johari2009} & Flow Control & System-wide & Full & Yes \\ 
\cite{Jain2010} & Flow Control & System-wide & Full & Yes \\ 
\cite{Kakhbod2012_TAC} & Flow Control & System-wide & Partial & Yes \\ 
\cite{Farhadi2019} & Joint Flow Control \& Multipath Routing & System-wide & Full & No \\ 
\cite{Kakhbod2012} & Power allocation \& Spectrum sharing & System-wide & Full & Yes \\ 
\cite{Bhattacharya2016} & Electric Vehicles & System-wide & Full & Yes \\ 
\cite{Sharma2012} & Networked Public Goods & Local & Full & Yes \\ 
\cite{Sinha2017_a} & Networked Public Goods & System-wide & Full & Yes \\ 
\cite{Sinha2014} & Network Utility Max & System-wide \& Linear & Full & Yes \\ 
\hline
\end{tabular}
\label{tab:Comm_table}
\end{table}

\subsubsection{VCG-based Mechanisms for Communications Networks}

``How can we allocate a fixed amount of an infinitely divisible resource among a finite set of strategic agents?" This is a classical problem in communication problems where pricing plays a key role \cite{Johari2009,falkner2000overview,shenker1990theoretical}. Under natural network constraints, and if our desirable outcome is to achieve efficiency, then our starting point is the \emph{Kelly mechanism} \cite{Kelly1997}. Briefly, the Kelly mechanism asks each agents to act as a bidder in an auction and announce a bid. Then the allocation of the single resource is conducted in proportion of the agents' bids. For each allocation, the social planner will receive an equal amount of tax from the agent's bid. In economics, we call this a ``market-clearing price." Such a price has many desirable characteristics as it ensures the bare-minimum communication (one-dimensional messages) and the social planner only needs to communicate back a single price per unit of resource. Both characteristics are ideal as they ensure the mechanism is practical, easily implementable for large-scale systems. The usefulness of the Kelly mechanism led to extensive research and the birth of \emph{scalar-parameterized mechanisms} (as the only communication between agents and the social planner is a scalar). Mathematically, the allocation function for agent $i \in \mathcal{I}$ in the Kelly mechanism is given by $x_i = \frac{c \cdot p_i}{\sum_{j \in \mathcal{I} \setminus \{i\}} p_j}$, where $c \in \mathbb{N}$ is the capacity of a single resource.

Following a similar methodology, a special formulation of the VCG mechanism for an auction setting was introduced in \cite{Lazar1998,Semret1999}. The problem was for a single divisible good under two different scenarios. At first, Semret \cite{Semret1999} studied non-differentiable pseudo-utility functions between a social planner and multiple agents. In \cite{Lazar1998}, the case was that agents seek bundles of links of a communication network (e.g., route) and the each agent's utility function depends on the minimum allocation received along their route. The key approach here was that for each link of the network a completely new and different auction was run. Both words have been somewhat generalized for multiple divisible links \cite{Bitsaki2005,Maille2004}. A general convex VCG-based mechanism was introduced in \cite{Johari2009} following \cite{Yang2007} which just like in the Kelly mechanism only required one-dimensional bid signals from all agents. Following this work, Yang and Hajek \cite{Jain2010} proposed a VCG-based mechanism with agents reporting a two-dimensional bid, i.e., a per-unit price $\beta$ and a maximum quantity $d_i$ that agent $i \in \mathcal{I}$ is willing to pay for the resource. Thus, the valuation function takes the form of
    \begin{equation}\setcounter{equation}{14}
        \hat{v}_i(x_i) = \beta \cdot \min\{x_i, d_i\}.
    \end{equation}
The benefit of this mechanism is it provides the equilibrium of the induced game (auction in this case) results to an optimal solution of the utility maximization problem based on the reported by the agents utility functions. Each agent pays for their allocation exactly the externality they impose on the other agents by participating in this mechanism. Of course, other mechanisms have been investigated, explored, and developed in great length \cite{loiseau2013incentive,shakkottai2008price,ozdaglar2007incentives,menache2011game,acemoglu2022too}.

\subsection{Power Grid Systems}

Over the last few decades, considerable efforts have been made to decarbonize our society's increasing energy demands \cite{ParisAccords2015} and reassess the efficiency of the existing methods of producing, managing, and consuming electricity \cite{ketter2018information}. Since 1980s \cite{gellings1985concept,fahrioglu2000designing,ramanathan2008framework}, \emph{demand-size management} (DSM) programs have been the standard way to study power-grid systems, their efficiency, and recently, their transformation to smart grid systems thanks to automation. Based on the US Energy Information Administration department: ``Demand-side management (DSM) programs consist of the planning, implementing, and monitoring activities of electric utilities which are designed to encourage consumers to modify their level and pattern of electricity usage." Hence, it is no surprise that the next-generation smart power grids utilize information and communication technologies as well as improved computational and sensor capabilities. That is why smart power grid systems are excellent examples of cyber-physical systems characterized by an overlay of information, algorithms, and enhanced operational programs to generate, transmit, distribute, and use electricity \cite{samad2017controls}.

There are multiple methodologies for DSM programs, but depending on the application and modeling choices, we can say that there are four main categories: (i) \emph{energy efficiency}, (ii) price-based \emph{demand response}, (iii) incentive-based demand response, (iv) {market-based} models. For the purposes of this article, our focus will be in the last category, i.e., \emph{market-centric} grid control. The key architectural structure for such models is as follows: there are \emph{generators}, \emph{loads}, \emph{distribution grids}, and \emph{aggregators}, all playing important roles as participants in the market. There is an \emph{independent system operator} (ISO) who is tasked to manage all and any transactions for electricity in the market \cite{kirschen2003demand}. The goal of such model is to provide the ``right" incentives and set of rules in order the overall power generation matches the load at all times cost-effectively and in an efficient manner \cite{samad2017controls}. The main approach to achieve this goal is for the ISO to assume complete control of the market and by introducing an auction, derive the incentives (e.g., electricity payments) for the efficient power and energy allocation under different scenarios. For example, we can consider different regulations and constraints in the production or distribution of electricity. Such modeling has been studied extensively the last ten years, as efforts to provide more sustainable consumption and production of electricity to homes from factories have been a top priority \cite{maharjan2013dependable,alpcan2004game,wang2014game,konstantakopoulos2016smart,amin2015game}. Two key challenges that have been studied in the literature are: (i) how can we design the right monetary payments in selling electricity to consumers in day-ahead or real-time settings? (ii) How can we ensure the efficient and balanced production and distribution of electricity as well as the control of the electrical voltage frequency? In parallel to market-based approaches, game-theoretic models have also been studied and explored in an effort to understand the strategic interactions between producers (e.g., production centers or factories) and consumers (e.g., households, buildings) \cite{coogan2013energy,zhou2017eliciting,zhou2017hedging,yang2015indirect,ma2016incentivizing}. This means that both producers and consumers are assumed selfish (rational and intelligent), and thus, make decisions (how much electricity to produce, how much electricity to consume) according to their own individual self-interest \cite{li2020transactive}.

Suppose our goal is to develop a pricing mechanism for a DSM program, in which we want to encourage the efficient energy consumption among consumers. If we adopt the mechanism design approach, then we set a system-wide socially-efficient objective that we want to achieve (this could be properties such as truthfulness, efficiency, and budget balanced). What if though all household appliances that use electricity from our consumers are jointly scheduled? One way to tackle this issue (since it causes severe computational complexity) is to use the technique \emph{consumer-level control}, in which we attempt to determine the total electricity consumption in each time step. Then we schedule enough production/distribution for the consumer's appliances to operate at the ``desired" electricity usage levels \cite{xiong2011smart,alizadeh2011information}. However, additional information from the consumers might be necessary and as it have been shown \cite{Samadi2012}, this information is private and consumers have no reason to report it truthfully. Hence, mechanism design can be used to provide a solution to this information elicitation problem by the appropriate design of incentives \cite{samadi2010optimal}. The literature can be categorized as follows: (i) auction-based mechanisms (mostly extensions of the VCG), (ii) market-based mechanisms \cite{Wilson2002,Cramton2003,Cramton2017,Bose2018,Karaca2020,Khazaei2019,Silva2001}, and (iii) indirect mechanisms for bidding and pricing models. In the next subsection, we review a general market-based framework of a smart grid system.

\subsubsection{Electricity Markets}

In this subsection, we offer a general formulation of an electricity market followed by its natural extension to an optimal mechanism design problem. Our exposition follows the work in \cite{Heymann2017_Chapter}.

We consider a smart power grid system consisted of agents (e.g., producers, consumers), a power grid network, and the electricity demand that comes from the agents. In this general framework, we also consider that the information is uncertain. Electricity production incurs a costs that needs to be covered by the producers. To capture this, we introduce a \emph{production cost function}. Thus, it is natural to expect all producers to be selfish (non-altruistic), and so their goal is to cover at the very least their cost when selling the produced electricity. In general, there are two ways we can model this problem: (i) use of \emph{bid function}, which directly maps electricity quantity into a payment; (ii) use of \emph{supply function}, which directly maps payments to the produced quantity. As we discussed earlier, there is an ISO, who aggregates the demand size. In mechanism design terms, ISO plays the role of a social planner and is tasked to elicit any private information, gather the bids and payments from the agents, and allocate the electricity that has been produced. Next, consider that $i \in \mathcal{I}$ refers to a node (some producer) of an arbitrary network $\mathcal{G}$, where $\mathcal{I}$ is the set of nodes. Each $i \in \mathcal{I}$ may produce $q_i \in \mathbb{R}$ quantity of electricity. As this is an interconnected power grid system, we say that the quantity of electricity that is sent from some node $i \in \mathcal{I}$ to node $j \in \mathcal{I}$ is denoted by $h_{i, j} \in \mathbb{R}$. At each node $i \in \mathcal{I}$, we assume that there is demand for electricity $d_i \in \mathbb{R}$. Each node is asked by the ISO to report a bid denoted by $b_i(q_i) \in \mathbb{R}$, and at the same also report the maximum quantity $\bar{q}_i$ node $i \in \mathcal{I}$ can produce. Next, we can add constraints to our problem to capture the specifics of the power grid network. We have $\mathbf{h} = (h_{i, j})_{i, j \in \mathcal{I}} \in \mathcal{H}$ such that $h_{i, j} \leq h_{i, j} ^ {\max}$. The goal of the ISO is to derive the allocation of electricity such that the total cost of production is minimized for the nodes (or producers), respect all constraints, and ensure the supply of electricity is at least greater than the demand (that way the power grid system can safely meet the demand) \cite{Heymann2017_Chapter}. Next, we add a nodal constraint of the form
    \begin{equation}
        q_i + \sum_{j \in \mathcal{I}} \left( h_{j, i} - h_{i, j} \right) - \sum_{j \in \mathcal{I}} \frac{c_{j, i} (h_{j, i}) + c_{i, j} (h_{i, j})}{2} \geq d_i,
    \end{equation}
where $c_{i, j}(\mathbf{h})$ represents the loss from the distribution of electricity with quantity $\mathbf{h}$ from node $i \in \mathcal{I}$ to node $j \in \mathcal{I}$. With these two constraints, we are ready to state the allocation of electricity optimization problem: for each $h \in \mathcal{H}$, we have
    \begin{gather}
        \min_{(q_i)_{i \in \mathcal{I}}, \mathbf{h}} \sum_{i \in \mathcal{I}} b_i(q_i) \\
        \text{subject to: } q_i + \sum_{j \in \mathcal{I}} 
        \left( h_{j, i} - h_{i, j} \right) - \sum_{j \in \mathcal{I}} \frac{c_{j, i} (h_{j, i}) + c_{i, j} (h_{i, j})}{2} \geq d_i, \\
        q_i \in [0, \bar{q}_i], \\
        h_{i, j} \geq 0.
    \end{gather}
We consider that there exists a probability distribution $f_i$ over a set of potential production cost functions. Each node $i \in \mathcal{I}$ is characterized by a type, say $\theta_i$, which represents a production cost function that depends on the quantity $q_i$. So, based on the mechanism design expectation that no node $i \in \mathcal{I}$ will report their true production cost function $\theta_i$ truthfully, we need to formulate an optimization problem and design appropriate incentives in order to have efficient production and distribution of electricity in the power grid network. We have
    \begin{gather}
        \min_{q_j, h_{i, k}, p_j} \sum_{j \in \mathcal{I}} \mathbb{E} [p_j(\theta)] \\
        \text{subject to: } q_j(\theta) + \sum_{i \in \mathcal{I}} \left( h_{i, j} - h_{j, i} \right) - \sum_{i \in \mathcal{I}} \frac{c_{i, j} (h_{i, j}) + c_{j, i} (h_{j, i})}{2} \geq d_j, \\
        \mathbb{E} [p_j(\theta)] - \theta_j(q_j(\theta)) \geq \mathbb{E} [p_j(b)] - \theta_j(q_j(b)), \\
        \mathbb{E} [p_j(\theta)] - \theta_j(q_j(\theta)) \geq 0, \\
        h_{i, j}, p_j \geq 0,
    \end{gather}
where $\theta = (\theta)_{i \in \mathcal{I}}$ is the profile of production cost functions. To find the solution of the above optimization problem, we need a mechanism with allocation and payment rules $(q, p)$ that minimize the expected payments of all nodes. Some key approaches to find the optimal $(q, p)$ are to analyze first a NE based on \emph{Bayesian Bertrand games} \cite{escobar2013efficiency}, study the quadratic externalities of the system or study the Walrasian equilibrium and replicate the techniques used in wholesale electricity auctions \cite{escobar2008equilibrium,escobar2010monopolistic}. Of course, there are many other techniques for this problem, i.e., stochastic market mechanisms \cite{Abbaspourtorbati2015} and energy-reserve co-optimized markets \cite{Xu2017}.

\subsection{Transportation Systems}

Commuters in big metropolitan areas have continuously experienced the frustration of congestion and traffic jams \cite{Varaiya1993}. Several studies have shown the benefits of \emph{emerging mobility systems} (e.g., ride-hailing, on-demand mobility services, shared vehicles, self-driving cars) in reducing energy and alleviating traffic congestion in a number of different transportation scenarios \cite{Papageorgiou2002,malikopoulos2019ACC,Ntousakis:2016aa,chalaki2020hysteretic,zhao2019CCTA-2,zardini2020,chalaki2020TCST,Malikopoulos2020,mahbub2020Automatica-2,chalaki2021Reseq}. Some recent and comprehensive surveys on the methodologies and techniques used in smart mobility-on-demand systems, see \cite{Malikopoulos2016a,guanetti2018,Zardini2022}.

One of the standard approaches to alleviate congestion in a transportation system has been the management of demand size due to the shortage of space availability and scarce economic resources by imposing an appropriate \emph{congestion pricing scheme} \cite{Pigou2013,Ferguson2021}. Such an approach focuses primarily on intelligent and scalable traffic routing, in which the objective is to guide and coordinate decision makers in choosing routes that leads to optimizing the routing decisions in a transportation network \cite{Korilis1997,Silva2013}. Due to the economic nature of congestion in transportation, auctioning has been proposed \cite{iwanowski2003} to create a market of tolls in a network of roads. This led to a spark of studies in auctioning techniques \cite{markose2007,teodorovic2008,vasirani2011,carlino2013,zhou2014,collins2015,isukapati2015,raphael2015,olarte2018}. Game theory has been one of the standard tools that can help us investigate the impact of selfish routing on efficiency and congestion \cite{Krichene2015,Krichene2016,Lam2016,Marden2015,Hota2020,fisac2019hierarchical,dong2015differential}. By adopting a game-theoretic approach, advanced systems have been proposed to assign decision makers concrete routes or minimize travel time and study a Nash equilibrium (NE) under different congestion pricing mechanisms \cite{Brown2018,Chandan2019,Salazar2019,Chremos2020SocialDilemma,Chremos2020MechanismDesign,chremos2020SharedMobility,chremos2020MobilityMarket,chremos2021MobilityGame,chremos2022Prospect,chremos2022Equity}. On the other hand, this approach imposes a few important limitations. First, the implementability of auction-based tolling on highways is not straightforward due to the dynamic and fast-changing nature of emerging mobility systems \cite{chalaki2021CSM}. Second, it is also uncertain how the public (e.g., travelers, drivers) will respond with respect to toll roads in an auction setting. Understanding the travelers' and drivers' interests (willingness-to-pay, value of time), and the impacts on different sociodemographic groups become imperative for the efficient design of transportation system utilization.

\subsubsection{Routing/Congestion Games}

We start this subsection with a motivational example. Suppose we have a simple transportation network $\mathcal{G} = (\mathcal{V}, \mathcal{E})$ with two routes $A$ and $B$ (one shorter than the other), where $\mathcal{V}$ is the set of nodes and $\mathcal{E}$ is the set of edges/roads. The agents (in this case, the drivers) start at origin $o \in \mathcal{V}$ and need to choose either route $A$ or $B$ to reach a final destination $d \in \mathcal{V}$. If all drivers choose the shortest route, then naturally congestion will occur and all driver will experience travel delays. So, the goal here is to find the best possible coordination of traffic through the different routes from $o$ to $d$. We can model this as a game, in which the drivers play against each other and have two possible actions (route $A$ or route $B$), thus leading to four possible outcomes. \emph{Which one is the best equilibrium?} The answer to this question has been studied extensively the last twenty years in the form of noncooperative \emph{routing games}, in which selfish agents compete for the best route in traffic of a transportation network \cite{roughgarden2002selfish}. Most interestingly, the generalization of routing games was developed by economist Robert W. Rosenthal in 1973 \cite{Rosenthal1973,rosenthal1973network}, in which the theoretical framework of \emph{congestion games} was introduced as noncooperative games of competing agents who have strategies subsets of resources. The agents' utility then depends only on the number of other agents who have chosen the same or overlapping strategy \cite{Correa2004,Krichene2015}. Moreover, the final cost of an agent can be computed as the sum of the costs of the strategy's elements (route, origin-destination pair). Mathematically, following the formulation of congestion games in \cite{Nisan2007} closely, we have $n \in \mathbb{N}$ agents with set $\mathcal{I} = \{1, \dots, n\}$ and a network $\mathcal{G}$. The set of strategies of each agent $i \in \mathcal{I}$ is a set of subsets of the set of edges $\mathcal{E}$. We call these \emph{paths} or \emph{routes}. For each edge $e \in \mathcal{E}$, there exists a \emph{congestion function} $c_e : \mathcal{I} \to \mathbb{R}_{\geq 0}$. Additionally, denote by $\mathcal{P} = \{(\mathcal{P}_i, \dots, \mathcal{P}_n)\}$ the set of paths (i.e., the strategies). The congestion of edge $e \in \mathcal{E}$ is some function $\ell(\mathcal{P})$ that can measure how many agents utilize one particular edge of agent $i \in \mathcal{I}$. Then agent $i$'s utility is given as
    \begin{equation}
        u_i = \sum_{e \in \mathcal{P}_i} c_e(\ell(\mathcal{P})).
    \end{equation}
If the set of agents is infinite, then the game is \emph{nonatomic} \cite{Schmeidler1973}. This notion, together with Rosenthal's congestion game framework allowed the formulation of general traffic networked games with an infinite number of agents, each with a negligible effect on the system's overall performance. Hence, nonatomic congestion games (and their special case: routing games) have been the standard model of transportation systems, communication routing networks, and computer science \cite{fabrikant2004complexity,Milchtaich2000,Milchtaich2004,Chau2003,Roughgarden2004,jalota2022credit,thai2016negative,krichene2014stackelberg,drighes2014stability,krichene2017stackelberg}. Moreover, this framework of congestion games has been proven to be quite flexible for a number of models and applications (see  \cite{roughgarden2002selfish}).

Another example to showcase the flexibility and the theoretical prowess of nonatomic congestion games is the traffic routing model \cite{Wardrop1952}. Agents are characterized as different origin-destination pairs in a transportation network of which its edges are the resources. The core assumption here is that agents compete for the limited (easily congested) resources. Thus, the strategies of the agents are represented as the possible paths between an origin and a destination. To each path, a cost is associated and it models the travel latency or travel delays based on the traffic along the path. The goal in this model is to find an equilibrium as closest to the \emph{social optimum}: a multicommodity flow of minimum total delay. In \cite{Wardrop1952}, the \emph{Wardrop equilibrium} was introduced as the collective strategies of all agents in network with the shortest paths. The significance behind the Wardrop traffic model under the congestion game framework is that it allows us to study the efficiency of equilibria. Both NE and Wardrop equilibria fail to minimize the social cost, thus are overtly inefficient. Thus, to improve the game-theoretic analysis, in 1999, Koutsoupias and Papadimitriou \cite{Koutsoupias1999,Papadimitriou2001} proposed the inefficiency of equilibria to be studied through the lens of worst-case analysis. In their seminal work, they introduced the ``Price of Anarchy," which represents the ratio of the worst social cost evaluated at a NE to the cost of an optimal solution \cite{Correa2004}. This key notion was extensively studied in transportation, communication, and computer science problems (e.g., selfish routing) \cite{Roughgarden2002,Roughgarden2004_b,Roughgarden2003,Chau2003,Correa2004,Perakis2007}.

\subsubsection{Auction-based Approaches for Intelligent Transportation Systems}

Traffic congestion persists to be one key challenge for the next-generation smart cities. The cost incurred by travel delays, traffic accidents, and fuel consumption has been estimated to be in the billions of US dollars per capita annually. That is why over the last twenty years \emph{Intelligent Transportation Systems} (ITS) have been introduced to provide solutions and make transportation in urban areas, as well as highways, safer, efficient, and convenient for the travelers and drivers. ITS are a multidisciplinary field as they incorporate multiple technologies such as wireless communication, navigation, sensing, and computing technologies \cite{zhou2013two}. For example, ITS have been applied in vehicle navigation, traffic signal control, emergency notification, and collision avoidance system. Naturally, communication is vital in such systems, and there are two main types of communication in ITS: \emph{vehicle-to-vehicle} and \emph{vehicle-to-infrastructure} (V2I) communication \cite{milanes2012intelligent}, in which advanced wireless communication allows vehicles to communicate crucial traffic information (e.g., speed, position, acceleration, traffic conditions, congestion, traffic warnings) and vehicles to infrastructure (some central authority/coordinator).

As we discussed earlier in this section, a key approach to the analysis of ITS can be based on auctions. Vickrey's work \cite{Vickrey1969} established \emph{congestion pricing} as a way to control congestion efficiently by requiring the travelers/drivers in the transportation system to pay tolls (as a function of the existing congestion, time, location, vehicle type) \cite{lindseytraffic}. Dynamic pricing has also been introduced, notably \cite{friesz2007computable,toh1997curbing}. In contrast to congestion pricing, an auction-based technique focuses on establishing an auction with competing vehicles reporting bids for ``time-slots" in order for them to travel in high-demand urban areas during peak hours \cite{teodorovic2008}. Alternatively, taking advantage of V2I technologies, a \emph{combinatorial auction} can be formulated to determine the right toll prices for vehicles \cite{zhou2014} (in such auctions, buyers compete with each other and bid to acquire multiple different but related goods). Based on \cite{teodorovic2008}, we have the following framework: there is a set of agents $\mathcal{I}$, where each agent $i \in \mathcal{I}$ makes a total of $m_i \in \mathbb{N}$ bids to enter the high-demand urban area. Time is modeled as a set of discrete steps, i.e., $\mathcal{T} = \{1, 2, \dots, T\}$, $T \in \mathbb{N}$. Each time interval represents the duration of an agent's stay in the area. So, each agent $i \in \mathcal{I}$ bids a monetary payment of value $p_{i j} \in \mathbb{R}$ for the right to visit the urban area a number of $c_{i j} \in \mathbb{N}$ times. Next, we have the following binary variables:
    \begin{equation}
        d_{i j}(t) =
            \begin{cases}
                1, & \text{if } c_{i j} \text{ consists of time interval } t, \\
                0, & \text{otherwise}.
            \end{cases}
    \end{equation}
In addition, we have
    \begin{equation}
        x_{i j}(t) =
            \begin{cases}
                1, & \text{if } c_{i j} \text{ is accepted}, \\
                0, & \text{otherwise}.
            \end{cases}
    \end{equation}
Thus, the optimization problem is formulated as follows:
    \begin{gather}
        \max_{x} \sum_{i \in \mathcal{I}} \sum_{j = 1} ^ {m_i} p_{i j} \cdot x_{i j}, \\
        \text{subject to: } \sum_{i \in \mathcal{I}} \sum_{j = 1} ^ {m_i} d_{i j}(t) \cdot x_{i j}(t) \leq D_{\max}, \quad \forall t \in \mathcal{T}, \\
        \sum_{j = 1} ^ {m_i} x_{i j} \leq 1, \quad \forall i \in \mathcal{I},
    \end{gather}
where $D_{\max}$ is the maximum number of vehicles that are allowed to be in the specific urban area. This optimization problem is rather difficult to solve as it is combinatorial and depending on its size, it can be almost impossible to solve in finite time. However, the auction-based congestion pricing \cite{teodorovic2008} can provide good-enough solutions. In \cite{zhou2014}, a similar yet improved combinatorial optimization problem is proposed and using VCG-based incentives provides an efficient solution in the alleviation of traffic.

More recently, auctions and mechanism design have been used in ridesharing \cite{karamanis2020assignment,Bian2020} and autonomous vehicles public transportation problems \cite{lam2015combinatorial}. We summarize some the latest papers in this area in Table \ref{tab:transportation_table}.

\begin{table}[ht]
    \centering
    \caption{A summary of auction-based mechanisms for transportation systems.}
        \begin{tabular}{|p{2.2cm}|p{4cm}|p{3cm}|p{4cm}|}
            \hline
                \multicolumn{4}{|c|}{Auction-based mechanisms for transportation systems} \\
            \hline
                Reference & Model & Auction Type & Scenarios \\
            \hline
\cite{zhang2015discounted} & Dynamic Ride-sharing & Combinatorial double auction & One-to-one assignment \\
\cite{kleiner2011mechanism} & Dynamic Ride-sharing & Vickrey & One-to-one assignment with detours \\
\cite{zhao2014incentives} & Dynamic Ride-sharing & Combinatorial double auction & One-to-one assignment with detours \\
\cite{lam2015combinatorial} & Dial-a-Ride problem & VCG & One-to-many \\
\cite{asghari2017line} & Dial-a-Ride problem & Combinatorial double auction & One-to-many \\
\cite{james2018double,asghari2016price} & Dial-a-Ride problem & VCG & One-to-one assignment with detours \\
            \hline
        \end{tabular}
    \label{tab:transportation_table}
\end{table}

\subsection{Security Systems}


It is well-known that cyber-defense remains a top priority for many organizations across different sectors. Part of this is because cyber-attacks are continuous and have the ability to bring down vital systems. Although there are many different angles to study security in a system of multiple agents, one approach is assume security is an economic good or resource and adopt game-theoretic approaches to study how we can find the best possible processes (via the appropriate design of incentives) for an efficient security investment \cite{alpcan2010network,manshaei2013game}. The main assumption in this approach is that the interactions between strategic agents in a system in which security is of crucial importance, can constitute a game. The objective of such a game is the provision of security in the means of investments by the agents. Thus, we say that security is a public good. This approach was first introduced by \cite{Kunreuther2003} as a way for a ``security game" to study airlines' baggage checking systems and what incentives are best. In parallel, Varian \cite{Varian2004} studied security games for the reliability of computer systems. Afterward security games were extensively studied in different fields and applications \cite{Kunreuther2003,Varian2004,Grossklags2008,Grossklags2010,Lelarge2009,Jiang2010}. There two surveys on this topic \cite{Manshaei2013,Laszka2012}.

\subsubsection{The Design of Security Games}

What is a security game? In general, interconnected systems of multiple agents who each depend on each other can be vulnerable to attacks by outsiders and external forces. Thus, any agent is encouraged to invest in the system's security measures not only to protect themselves but also to protect the other agents. Interconnectedness implies a positive externality. Consequently, the provision of security can be modeled naturally as the provision of a \emph{public good} (non-rivalrous commodity) \cite{Mas_Colell1995}. Thus, a security game models the strategic interactions of agents in a security system where each is asked to invest their own resources to secure the system from external attacks. Mathematically, we consider a network of $n \in \mathbb{N}$ agents (e.g., networked computer servers, corporate divisions, self-driving vehicles on a highway). The utilities of all agents are interdependent, and so $v_i(\mathbf{x})$. Each agent is assumed to possess a finite amount of resources $w_i \in \mathbb{N}$ available for investment, and we say that if an attack is successful against agent $i$'s resources then a loss $\ell_i \in (0, w_i]$ may be imposed to agent $i \in \mathcal{I}$. The agent is allowed to protect their resources, hence agent $i \in \mathcal{I}$ may invest $x_i \in \mathbb{R}_{\geq 0}$. However, this investment comes with a cost, represented by a general function $c_i : \mathbb{R}_{\geq 0} \to \mathbb{R}_{\geq 0}$ evaluated at the individual amount of investment $x_i$. We denote the vector of all agents' investments by $\mathbf{x} = \{ x_1, \dots, x_n \} = \mathcal{X}$. Then to capture how likely an attack may be successful or not, we introduce a risk function $r_i : \mathbb{R}_{\geq 0} ^ n \to [0, 1]$. Let $\mathbf{x}_{- i} = \{ x_1, \dots, x_{i - 1}, x_{i + 1}, \dots, x_n \}$ capture the interdependence of all other agents' security. Then, the utility of an agent $i \in \mathcal{I}$ is of the form
    \begin{equation}
        v_i(\mathbf{x}) = w_i - \ell_i \cdot r_i(\mathbf{x}) - c_i(x_i).
    \end{equation}
Hence, the security game is given by the tuple $\langle \mathcal{I}, \mathcal{X}, (v_i)_{i \in \mathcal{I}} \rangle$ under a complete or incomplete information setting (depending on the problem).

The transformation of a security game to a mechanism design problem is straightforward. For a solution concept, we adopt the NE. Formally, we have
    \begin{equation}
        \tilde{x}_i = \arg \max_{x \geq 0} v_i(x_i, x_{- i}).
    \end{equation}
For completeness, the socially-optimal solution profile $\mathbf{x} ^ *$ can be computed as follows
    \begin{equation}
        \mathbf{x} ^ * = \arg \max_{x \geq 0} \sum_{i \in \mathcal{I}} v_j(\mathbf{x}).
    \end{equation}
Next, the total interdependent utility of agent $i \in \mathcal{I}$ is given by
    \begin{equation}
        u_i(\mathbf{x}, p_i) = v_i(\mathbf{x}) - p_i,
    \end{equation}
where $p_i$ represents a payment/tax for agent $i \in \mathcal{I}$ when the investment profile is $\mathbf{x}$. Naturally, in a security system, it is imperative to design the right payments for all agents in order to incentivize voluntary participation. One approach for this is to introduce the notion of \emph{exit equilibrium}: this is an equilibrium that takes into account the ``external strategy" of an agent from the mechanism for the provision of a non-excludable goods, such as security and investment. At an exit equilibrium, an agent may unilaterally opt out of the mechanism and decide to adopt their best-response against the other agents who have chosen to participate in the mechanism \cite{naghizadeh2016opting}. Formally, we have
    \begin{align}
        \hat{\mathbf{x}}_{- i} ^ i & = \arg \max_{\mathbf{x}_{- i} \geq 0} \sum_{j \neq i} v_j(\mathbf{x_{- i}}, \hat{\mathbf{x}}_i ^ i), \\
        \hat{\mathbf{x}}_i ^ i & = \arg \max_{x_i \geq 0} v_i(\hat{\mathbf{x}}_{- i} ^ i, x_i).
    \end{align}
Most notably, although this is a clever notion of an exit equilibrium, Naghizadeh and  Liu \cite{naghizadeh2016opting} showed an impossibility result for mechanisms that induce a security game with  social optimality, weak budget balance, and voluntary participation. Most interestingly, two main approaches, i.e., the VCG mechanism and the \emph{externality mechanism} \cite{hurwicz1979outcome,sharma2011game} that have been quite successful in other applications fail to circumvent this theoretical obstacle. For the purpose of this article, we focus on the latter only. 

The externality mechanism aims to redistribute the wealth that collects from the agents and ensure strong budget balance (all payments equal to zero). As it was shown in \cite{sharma2011game}, this mechanisms induces a game with social optimality, voluntary participation, and maintains a balanced budget for cellular networks (which is an excludable public good). The payment function
    \begin{equation}
        t_i(\mathbf{x} ^ *) = - \sum_{j \in \mathbb{I}} x_j ^ * \ell_i \frac{\partial r_i}{\partial x_j} (\mathbf{x} ^ *) - x_i ^ * \frac{\partial c_i}{\partial x_i} (x_i ^ *).
    \end{equation}
These taxes fail to induce voluntary participation for all agents, and thus the system will be significantly vulnerable to attacks as its security cannot be guaranteed.

\subsubsection{Security in Communication Systems}

\begin{table}[ht]
    \centering
    \caption{A snapshot of papers on security of communication systems.}
        \begin{tabular}{|p{4cm}|p{4cm}|p{3cm}|p{3cm}|}
            \hline
                \multicolumn{4}{|c|}{Auction-based mechanisms for security in communication systems.} \\
            \hline
                Auction Type & Characteristics & Solution Concept & Scenario \\
            \hline
English auction \cite{Krishna2009} & Winner pays second highest price & NE & Information integrity \\
Dutch auction \cite{Krishna2009} & Winner pays final price & NE & Black-hole attack \\
First-price sealed-bid auction \cite{lucking2000vickrey} & Winner pays highest price & NE & Privacy or faked-sensing attack \\
Vickrey auction \cite{lucking2000vickrey} & Winner pays second highest price & NE & User collusion/bid-rigging \\
VCG auction \cite{ausubel2006lovely} & Vickrey auction with multiple goods & BNE & Eavesdropping attack \\
Share auction \cite{huang2007auction} & Matching of goods based on ratio of buyers' bids & NE & Distributed Denial-of-Service attack \\
Ascending-clock auction \cite{zhang2016ascending} & Increase price until demand equals supply & Walrasian equilibrium & False-name bids \\
Double auction \cite{friedman2018double} & Matching between sellers' demands and buyers' bids & Market equilibrium & Privacy \\
            \hline
        \end{tabular}
    \label{tab:security_table}
\end{table}

Security has been extensively studied in communication systems since the 1980s. With the advancement of communication technology such as wireless networks, communication between systems, computers, phones has never been as easy and quick. However, such systems admit vulnerabilities and since they play a key role in the exchange of information, attacks are continuous and frequent. So, in this subsection, we briefly focus on auction-based mechanisms that offer solutions for an array of communication scenarios and the solutions we can derive.

Auctions have been used to model the economic-in-nature interactions between agents of a communication system that need to invest resources for their security and wellbeing of the system (e.g., prevent external attacks). Examples of auctions used to model security include \emph{English auctions}, \emph{Dutch auctions}, \emph{Sealed-bid auctions}, \emph{Double auctions}, \emph{Share auctions}, and \emph{Ascending-clock auctions}.

Both English and Dutch auctions are multiple-round auctions and allow buyers to exchange information (their bid) to other buyers. In economics, this type of disclosure in auctions is called \emph{open-outcry}. In addition, the auctioneer in English auctions introduces a bid and keeps increasing it at each round, while in Dutch auctions, the opposite process occurs (high price decreasing until accepted). Thanks to the simplicity of the auction rules, such auctions have been used to defend and identify ``malicious" agents in a system \cite{reidt2009fable}. Sealed-bid auctions, as their name suggest, are characterized by their ensured privacy as buyers do not disclose their bids to anyone (examples are the first-price auction, Vickrey auction, and the VCG auction). The specifics of each auction have been summarized in Table \ref{tab:security_table} together with recent key applications \cite{luong2017applications}. For example, the VCG auction extends the Vickrey auction with multiple goods and sets the right rules to determine the winner (i.e., the buyer who gets the auctioned good). In particular, the VCG auction awards the good to the buyer, who then pays the second-highest bid. This is of extreme importance as it establishes the following principle: bid your true valuation and pay less than you expect. The VCG auction has been widely used in securing communication and wireless networks (e.g., to prevent agent misbehavior).

So far, the above-mentioned auctions are \emph{one-sided}, meaning there is one auctioneer who tries to sell one or more goods to a finite number of buyers. In a double auction there is still an auctioneer who manages an auction between multiple buyers, who each submit bids at the same time, and sellers, who each submit payment demands for their goods \cite{friedman2018double}. For example, an auctioneer can match the buyers and sellers as follows: list the bids in a descending order and the payment demands in an ascending order. Denote by $p_m ^ {\text{seller}}$, $p_m ^ {\text{buyer}}$ for the payment demand of a seller and the bid of a buyer, respectively. Then the auctioneer can find the largest index $m \in \mathbb{N}$ for which we have
    \begin{equation}
        p_m ^ {\text{buyer}} \leq p_m ^ {\text{seller}}.
    \end{equation}
The next step is to set the final payment as
    \begin{equation}
        p ^ * = \frac{p_m ^ {\text{buyer}} + p_m ^ {\text{seller}}}{2}.
    \end{equation}
In economics, $p ^ *$ is called \emph{clearing price}. Using these prices, buyers get the good by paying $p ^ *$, and naturally, the seller receives $p ^ *$. This process can be repeated as many times as necessary and it stops once all sellers' goods have been matched or sold to the buyers.

Lastly, share auctions are a type of auctions that resemble a market and are used to model a resource allocation to multiple buyers. However, the resource needs to be divisible (e.g., bandwidth) \cite{huang2010game}. Share auctions are as follows: buyers submit to an auctioneer a bid (how much bandwidth they require) and then the auctioneer computes a payment that is proportional to the buyer's bid. For example, suppose there are $K \in \mathbb{N}$ source-destination pairs of friendly jammers (sellers of friendly-jamming power) and eavesdroppers in a wireless communication network. The goal is to improve the secrecy capacity of a source. A buyer submits a bid in a form of asking for friendly-jamming power, say $\pi_i \in \mathbb{R}$. Thus, we can allocate the friendly-jamming power at each source $i \in \mathcal{I}$, i.e.,
    \begin{equation}
        \pi_i = \frac{b_i \pi_{\max}}{\beta + \sum_{k = 1} ^ {K} b_k},
    \end{equation}
where $\pi_{\max}$ is the maximum possible power, $b_i$ is the source $i$'s bid, and $\beta \in \mathbb{R}$. finally, source $i \in \mathcal{I}$ pays $p_i = \lambda \pi_i$, where $\lambda \in \mathbb{R}$ is a payment per unit of power. Such auctions can be used to find the optimal bid between sources in the network and maximize the secrecy capacity change \cite{zhu2009improved}. A comprehensive survey and tutorial can be found in \cite{luong2016data,zhang2012auction}.

\section{Conclusion}

In this last section of our article, we offer a concluding discussion on the theory of mechanism design and its engineering applications. First, we discuss the theoretical and practical limitations of the theory, some of the most notable criticism and how to move forward. We pay extra attention in offering key open questions in mechanism design and propose two main future research directions that we, as authors, believe have the greatest potential. Lastly, we conclude this article with a few remarks.


Traditionally, mechanism design has focused mainly on quasilinear static settings under somewhat simple structures for the agents' types or utility functions. Some particular (unrealistic) strong assumptions have pervaded, making it difficult for mechanisms to be implemented in real-life problems. Whether the set of agents remains fixed and known to the social planner or the evolution of information is not considered, the ability to design mechanisms that offer a superior way to design incentives and induce efficiency across a system still remains a formidable challenge in mechanism design. For example, the VCG mechanism, although a widely-used mechanism, it is not frugal and depending on the application it might overtax agents (thus, impose a heavy tax burden). Information and communication, as it was envisioned by Leonid Hurwicz, relies on the universal existence and authority of a central authority (a social planner). Agents are expected to fully disclose their private information, raising privacy concerns as well as computational costs to manage the sheer amount of information. Another key critique of the theory is the computational intractability of certain mechanisms as many fail to provide an iterative learning process (alternatively called t\^{a}tonnement process), i.e., there exists an algorithm for the equilibrium to be attained by the agents and social planner.


In general, from an engineering standpoint, mechanism design is characterized by its trade-off between the design of optimal and efficient solutions that all agents will accept and realistic and system-wide properties such as simplicity, robustness, and computational trackability. As control problems are commonly dynamic, complex, and unpredictable \cite{sztipanovits2019science}, it still remains an open question to devise mechanisms are simple yet dynamic, robust, and trackable. At the same time, a key open question is to look at the intersection of mechanism design and machine learning allowing mechanisms with incentives that lead to efficient equilibria that can be learned in dynamic environments (i.e., extending the typical mechanism to address dynamic control problems). It is the authors' belief that engineering applications (e.g., communication networks, information systems, transportation networks) of large-scale systems, which are dynamic in nature, impose rather crucial challenges to the theoretical framework of mechanism design, and thus, inspire novel new mechanisms that will circumvent some of the limitations of the theory. The goal, of course, is to improve the applicability of these mechanisms in real-life problems and translate the usefulness of the theoretical insights into practice. Finally, game theory allows us to model the strategic interactions of systems consisted of multiple agents/players and who compete over resources. The theory of mechanism design allows us to adopt an objective-first approach and model the best possible game and its rules. As new developments and applications are continuous, it remains to be seen the next chapters of mechanism design in engineering and its true impact in solving the big problems.

\begin{tcolorbox}[parbox=false,enhanced,breakable]

\section{Sidebar: Further Reading}\label{SIDEBAR:MD_furtherreading}

There is a rich list of books in game theory, economic design, and mechanism design for the interested reader and especially for any first-year graduate students as a first resource to learn more about the field. The literature on game theory and its applications includes several excellent textbooks \cite{Fudenberg1991,Luce1989,Gibbons1994,Myerson2013,Osborne2009,Osborne1994}. Most of the the established theory of mechanism design can be divided into three categories: economics, auctions, and computer science (with general applications). The golden standard in microeconomics including rigorous introductions to game-theoretic notions and mechanism design is \cite{Mas_Colell1995}. A more focused on Nash-based mechanisms and matching markets textbook is \cite{Diamantaras2009}. A recent book \cite{Borgers2015} offers a more economics-inclined theory establishing the foundations of mechanism design. Excellent textbooks in implementation theory are \cite{Jackson2001} (with a great chapter in \cite{Osborne1994}). Robust mechanism design in economics has been mostly developed by Dirk Bergemann, who has compiled his decade work in book \cite{bergemann2012robust}. So far, most of the textbooks focus on implementation of mechanisms and the efficient design of incentives. A completely different approach, one that is much closer to the first seminal paper by Leonid Hurwicz, is the study of mechanisms under informational efficiency focusing mostly in the preservation of privacy \cite{Hurwicz2006} and the bounded size of the message space \cite{williams2008communication}. However, a rather engineering-accessible textbook that offers a theoretical framework of mechanisms through fundamental results from linear programming is \cite{vohra2011mechanism}. Mechanism design has been instrumental in auction design and many times the methodologies and techniques of the two theories are interchangeable. Thus, here are some great general textbooks \cite{Krishna2009,Milgrom2004}, and a survey on combinatorial auctions \cite{cramton2004combinatorial}. We end this part by offering excellent surveys from the economics literature \cite{roth2008have,cramton2013spectrum}. Of course, there have been numerous key contributions in mechanism design by computer scientists and in many graduate programs mechanism design is part of the curriculum. A well-known textbook that offer a survey of these contributions can be found in Chapters 9--16 of \cite{Nisan2007} and in general \cite{Shoham2008}. Additional resources in game-theoretic approaches and techniques of mechanism design can be found here (including dynamic incentive design) \cite{narahari2014game,mailath2018modeling,gershkov2014dynamic,glazer2016models}. Lastly, a recent publication that discusses the most recent developments, insights, and theoretical results in economic design is \cite{laslier2019future}.
\end{tcolorbox}

\section{Acknowledgements}
This work was supported by the National Science Foundation under Grants CNS-2149520 and CMMI-2219761. This support is gratefully acknowledged.

\bibliographystyle{Resources/IEEEtran.bst}
\bibliography{References/IDS_Publications_11232022.bib,References/References.bib}


\sidebars 









\newpage

\section{Author Biography}

{Ioannis Vasileios Chremos}
    (S'18) received the B.S. degree with honours of the First Class in Mathematics from the University of Glasgow, Glasgow, UK in 2017. He is working towards a Ph.D. in the program of Mechanical Engineering at the University of Delaware. His research interests lie broadly in emerging mobility systems, game theory, and mechanism design. Current research projects involve the study of the social impact of automation in emerging mobility systems with connected and automated vehicles and the holistic development of a sociotechnical system framework for smart cities using insights and techniques from behavioral economics and mechanism design. He is also interested in game-theoretic models of social networks and the study of different methodologies for the prevention of the spread of misinformation. He is a student member of the SIAM and ASME.

{Andreas A. Malikopoulos}
    (S'06--M'09--SM'17) received the Diploma in mechanical engineering from the National Technical University of Athens, Greece, in 2000. He received M.S. and Ph.D. degrees from the department of mechanical engineering at the University of Michigan, Ann Arbor, Michigan, USA, in 2004 and 2008, respectively. He is the Terri Connor Kelly and John Kelly Career Development Associate Professor in the Department of Mechanical Engineering at the University of Delaware (UD), the Director of the Information and Decision Science (IDS) Laboratory, and the Director of the Sociotechnical Systems Center. Before he joined UD, he was the Deputy Director and the Lead of the Sustainable Mobility Theme of the Urban Dynamics Institute at Oak Ridge National Laboratory, and a Senior Researcher with General Motors Global Research \& Development. His research spans several fields, including analysis, optimization, and control of cyber-physical systems; decentralized systems; stochastic scheduling and resource allocation problems; and learning in complex systems. The emphasis is on applications related to smart cities, emerging mobility systems, and sociotechnical systems. He has been an Associate Editor of the IEEE Transactions on Intelligent Vehicles and IEEE Transactions on Intelligent Transportation Systems from 2017 through 2020. He is currently an Associate Editor of Automatica and IEEE Transactions on Automatic Control. He is a member of SIAM, AAAS, and a Fellow of the ASME.

\end{document}